\documentclass[aps,prb,superscriptaddress,amsmath,amssymb,floatfix,twocolumn]{revtex4-1}
\usepackage{times}
\usepackage{graphicx}
\usepackage{subfigure}
\usepackage{color}
\usepackage{bm}
\usepackage{float}

\newcommand{\bsigma}{\boldsymbol{\sigma}}
\newcommand{\bk}{\mathbf{k}}
\newcommand{\dn}{\downarrow}
\newcommand{\pdag}{\phantom{\dag}}
\newcommand{\up}{\uparrow}

\begin{document}

\title{Quantum critical Kondo destruction in the Bose-Fermi Kondo model with a
local transverse field}
\author{Emilian Marius Nica}
\email[Corresponding author: ]{en5@rice.edu}
\affiliation{Department of Physics and Astronomy, Rice University, Houston,
Texas 77005}

\author{Kevin Ingersent}
\affiliation{Department of Physics, University of Florida, P.O.\ Box 118440,
Gainesville, Florida 32611}

\author{Jian-Xin Zhu}
\affiliation{Theoretical Division, Los Alamos National Laboratory, Los Alamos,
New Mexico 87545}

\author{Qimiao Si}
\affiliation{Department of Physics and Astronomy, Rice University, Houston,
Texas 77005}

\date{\today} 

\begin{abstract}
Recent studies of the global phase diagram of quantum-critical heavy-fermion
metals prompt consideration of the interplay between the Kondo interactions
and quantum fluctuations of the local moments alone.
Toward this goal, we study a Bose-Fermi Kondo model (BFKM) with Ising
anisotropy in the presence of a local transverse field that generates quantum fluctuations in the local-moment sector.
We apply the numerical renormalization-group method to the case of a sub-Ohmic
bosonic bath exponent and a constant conduction-electron density of states.
Starting in the Kondo phase at zero transverse-field, there is a smooth crossover with
increasing transverse field from a fully screened to a fully polarized impurity
spin. By contrast, if the system starts in its localized phase, then
increasing the transverse field causes a continuous, Kondo-destruction
transition into the partially polarized Kondo phase. The critical exponents at
this quantum phase transition exhibit hyperscaling and take essentially
the same values as those of the BFKM in zero transverse field.  The many-body
spectrum at criticality varies continuously with the bare transverse
field, indicating a line of critical points. We discuss implications of these
results for the global phase diagram of the Kondo lattice model.
\end{abstract}

\maketitle

\section{Introduction}\label{Sec:Intro}

Heavy fermions form a class of rare-earth based intermetallic compounds that has
attracted sustained attention.\cite{si_science10} 
Recent years have seen intensive effort, both in
theory and experiment, to understand the unusual properties exhibited
by these materials over a temperature range above a quantum critical point
(QCP).\cite{stewart,gegenwart_natphys08,HvL} 
The most typical cases involve a zero-temperature transition
from an antiferromagnetically ordered state to a paramagnetic heavy Fermi-liquid.
A particularly notable feature of the quantum-critical regime is the non-Fermi
liquid behavior, which has been observed in transport, thermodynamic, and other
properties in a number of compounds. Two fundamentally different classes
of quantum critical points have been proposed theoretically.
The spin-density-wave QCPs\cite{hertz} represent the quantum-mechanical extension
of the classical Landau-Ginzburg-Wilson framework, describing criticality
solely in terms of fluctuations of a magnetic order parameter. By contrast, the
locally critical picture\cite{lcqpt:01,lcqpt:03,colemanetal} is ``beyond-Landau''
in that it invokes the destruction of the heavy quasiparticles at the transition.
Such a Kondo destruction introduces new critical degrees of freedom beyond
order-parameter fluctuations. A microscopic theory of local quantum
criticality\cite{lcqpt:01,lcqpt:03} has been formulated in terms of extended
dynamical mean-field theory (EDMFT),\cite{SmithSi,Chitra:00} in which the Kondo
lattice model is mapped to an effective quantum impurity problem: the Bose-Fermi
Kondo model (BFKM) with self-consistently determined densities of states for
the fermionic conduction band and for the bosonic bath.

Experimental evidence for local quantum criticality has come from systematic
studies in several heavy-fermion materials. In YbRh$_2$Si$_2$ and CeRhIn$_5$, 
the large Fermi surface of the paramagnetic metal phase has been shown to
collapse at the antiferromagnetic QCP,\cite{paschen04,friedemann10,shishido}
providing direct evidence for a critical destruction of the Kondo effect. The
critical dynamical spin susceptibility at the QCP in Au-doped CeCu$_6$ departs
drastically from the predictions of the spin-density-wave picture, instead
satisfying $\omega/T$ scaling and displaying a fractional exponent in the
frequency and temperature dependence over a large region of the Brillouin
zone.\cite{schroder} Such scaling properties have been captured in EDMFT
calculations for Kondo lattice
models.\cite{lcqpt:01,lcqpt:03,Grempel.03,ZhuGrempelSi,Glossop.07,Zhu.07}

\begin{figure}[!h]
\includegraphics[width=1.0\columnwidth]{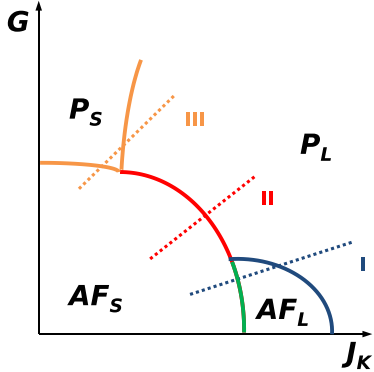}
\caption{(Color online)
Schematic $T=0$ phase diagram proposed for heavy-fermion metals described
by a Kondo lattice.\cite{Si:10+06} $J_K$ is the antiferromagnetic Kondo
coupling between local moments and conduction electrons, while $G$ parameterizes
increasing magnetic frustration or spatial dimensionality. Solid lines indicate
phase boundaries. Dashed lines labeled I, II, and III represent three different
types of route between the paramagnetic heavy Fermi-liquid phase having a large
Fermi surface (P$_{\textrm{L}}$)  and an antiferromagnetic Kondo-destroyed phase with a small 
Fermi surface (AF$_{\textrm{S}}$)}.

\label{figg1}
\end{figure}

More recently, the notion of Kondo destruction has been incorporated into
a global zero-temperature phase diagram for heavy fermions.\cite{Si:10+06}
The phase diagram, proposed for the Kondo lattice model, is shown in
Fig.\ \ref{figg1}, where the abscissa represents the Kondo exchange coupling between local moments and conduction electrons, while the ordinate G parameterizes quantum fluctuations of the local moments. Three different
sequences of quantum phase transitions can connect a Kondo-destroyed
antiferromagnet to the Kondo-entangled paramagnetic heavy-fermion state. This
diagram provides a framework for understanding not only the examples of
local quantum-critical behavior described above, but also the detachment of
the Kondo destruction transition from the antiferromagnetic transition as
evidenced\cite{11,12,13} in Ge- and Ir-doped YbRh$_2$Si$_2$ and in YbAgGe.
The global phase diagram has recently been employed to understand the dimensional
tuning of the quantum-critical behavior in heavy-fermion systems,\cite{custers12}
and has served as a motivation for a recent flurry of experiments on
heavy-fermion materials with geometrically frustrated
lattices.\cite{kim_aronson13,mun13,fritsch13,Khalyavin13}

%\begin{figure}[!htb]

The key feature of the global phase diagram for heavy fermions is the interplay
between quantum fluctuations related to the Kondo effect and those associated with
the local moments alone. Consider the Kondo lattice Hamiltonian
$H_{\mathrm{KL}} = H_c + H_I + H_K$, where $H_c$ and $H_I$ represent
a conduction-electron band and a lattice of exchange-coupled local moments,
respectively, and $H_K$ specifies the Kondo coupling between these two
sectors.
In situations where the local moments exhibit Ising (easy-axis) anisotropy,
quantum fluctuations of these moments can readily be generated through
application of a transverse magnetic field.
For example, in the stand-alone transverse-field Ising model described by
$H = H_I = \sum_{ij} I_{ij} S_i^z S_j^z + \Delta \sum_i S_i^x$,
the transverse field $\Delta$ introduces quantum fluctuations and sufficiently
large values of $\Delta$ suppress any magnetic order. In other words, $\Delta$
provides a realization of the parameter $G$ in Fig.~\ref{figg1}. Within the
EDMFT treatment of $H_{\mathrm{KL}}$, this interplay of fluctuations can
be described using a self-consistent BFKM with a static transverse magnetic
field. As a nontrivial first step towards solving this problem, we are led to
consider the impurity version of this problem without the imposition of
self-consistency.

This paper reports numerical renormalization-group (NRG) results for the BFKM in the
presence of a local transverse magnetic field $\Delta$. The conduction-electron
density of states is taken to be structureless, while the bosonic bath is
assumed to be characterized by a spectral exponent $s$ that takes a sub-Ohmic
value $0<s<1$. For $\Delta=0$, this model exhibits a Kondo-destruction QCP
separating a Kondo or strong-coupling phase, in which the impurity spin is completely quenched at temperature $T=0$, from a localized phase in
which spin-flip exchange scattering is suppressed and the impurity exhibits a local moment with a Curie magnetic susceptibility.\cite{Zhu:02,Zarand:02,
Glossop:05+07} For any combination of Kondo and bosonic couplings
that, at $\Delta=0$, places the system within the Kondo phase, increasing the
transverse field produces a smooth crossover to a fully polarized impurity spin
without the appearance of a quantum phase transistion. For couplings that 
localize the impurity spin in the absence of a transverse field,
increasing such a field eventually causes a continuous, Kondo-destruction
transition into the partially polarized Kondo phase. The critical exponents at
this quantum phase transition are found, for the particular case of bosonic
bath exponent $s=0.8$, to exhibit hyperscaling and to take essentially the
same values as those of the BFKM in zero transverse field. 
The critical NRG spectrum varies continuously with the bare transverse field,
indicating a line of critical points.

The remainder of the paper is organized as follows. Section II defines the
model and briefly describes the NRG solution method. Section III demonstrates the existence of a Kondo-destruction transition in the BFKM at particle-hole symmetry in the presence of a bosonic
bath characterized by a sub-Ohmic spectral exponent $s=0.8$.
In Sec. IV we interpret the critical spectra reached for different values of the transverse field as evidence for a line of
renormalization-group (RG) fixed points. The accuracy of our calculated critical exponents is addressed in an Appendix.

\section{Model, Qualitative Expectations and Solution Method}
\label{Sec:Model}

\subsection{Bose-Fermi Kondo model with a transverse field}

The Hamiltonian for the Ising-anisotropic BFKM with a transverse field can
be written
\begin{multline}
\label{H_BFKM}
H_{\mathrm{BFKM}}
= \sum_{\bk,\sigma} \epsilon_{\bk} c_{\bk\sigma }^{\dag}
   c_{\bk\sigma}^{\pdag} + \frac{J_K}{2} \, \mathbf{S} \cdot \!\!\!
   \sum_{\bk,\bk',\sigma,\sigma'} \!\!\! c_{\bk\sigma}^{\dag}
   \bm{\sigma}_{\sigma\sigma'} c_{\bk'\sigma'}^{\pdag} \\
+\sum_q \omega_q \phi_q^{\dag} \phi_q^{\pdag}
   + S_z \sum_q g_q \bigl( \phi_q^{\pdag}+\phi_{-q}^{\dag} \bigr)
   + \Delta S_x,
\end{multline}
where $\epsilon_{\bk}$ is the dispersion for a band of noninteracting conduction
electrons, $J_K>0$ is the antiferromagnetic Kondo coupling between a
spin-$\frac{1}{2}$ local moment $\mathbf{S}$ and the spin density of conduction
electrons at the impurity site, $g_q$ is the coupling of the impurity spin $z$ component to a bosonic degree of freedom with annihilation operator $\phi_q$,
and $\Delta\ge 0$ is the transverse field. 
We work in units where $g\mu_B=k_B=\hbar=1$.

Throughout the paper, we assume a featureless metallic conduction electron band
described by the density of states 
\begin{equation}
\label{rho:def}
\rho(\epsilon)
  = \sum_{\bk} \delta(\epsilon-\epsilon_{\bk})
  = \rho_0 \, \Theta(D-\vert \epsilon\vert) ,
\end{equation}
where $D$ is the half-bandwidth and $\Theta$ is the Heaviside step
function, allowing definition of a dimensionless exchange coupling $\rho_0 J_K$
between the band and the impurity spin. We denote by $T_{K}^{0}$ the Kondo
temperature associated with this exchange in the absence of any bosonic
coupling.
The bosonic spectral function is taken to be 
\begin{equation}
\label{B:def}
B\left(\omega \right)
\equiv \pi \sum_{q} g_q^2 \delta\bigl( \omega -\omega_q \bigr)
= B_0 \, \omega_0^{1-s} \omega^s \, \Theta(\omega) \, \Theta(\omega_0-\omega) ,
\end{equation}
where $\omega_0$ is a high-energy cutoff and $B_0$ is a dimensionless effective
coupling between the impurity spin and
the bosonic bath.
Previous studies\cite{Zhu:02,Glossop:05+07,27} have taken $g_q = g$, in
which case $B_0 = {(K_0g)}^{2}$ with $K_0$ being determined by the bath density
of states. Henceforth, we refer to $B_0$ as the effective coupling
to the dissipative bath. Values of the exponent $s < 1$,
$s=1$ and $s > 1$ correspond to sub-Ohmic, Ohmic and super-Ohmic baths,
respectively.

\subsection{Qualitative Considerations}\label{Sec:Qualitative}

\begin{figure}[!]
\includegraphics[width=1.1\columnwidth]{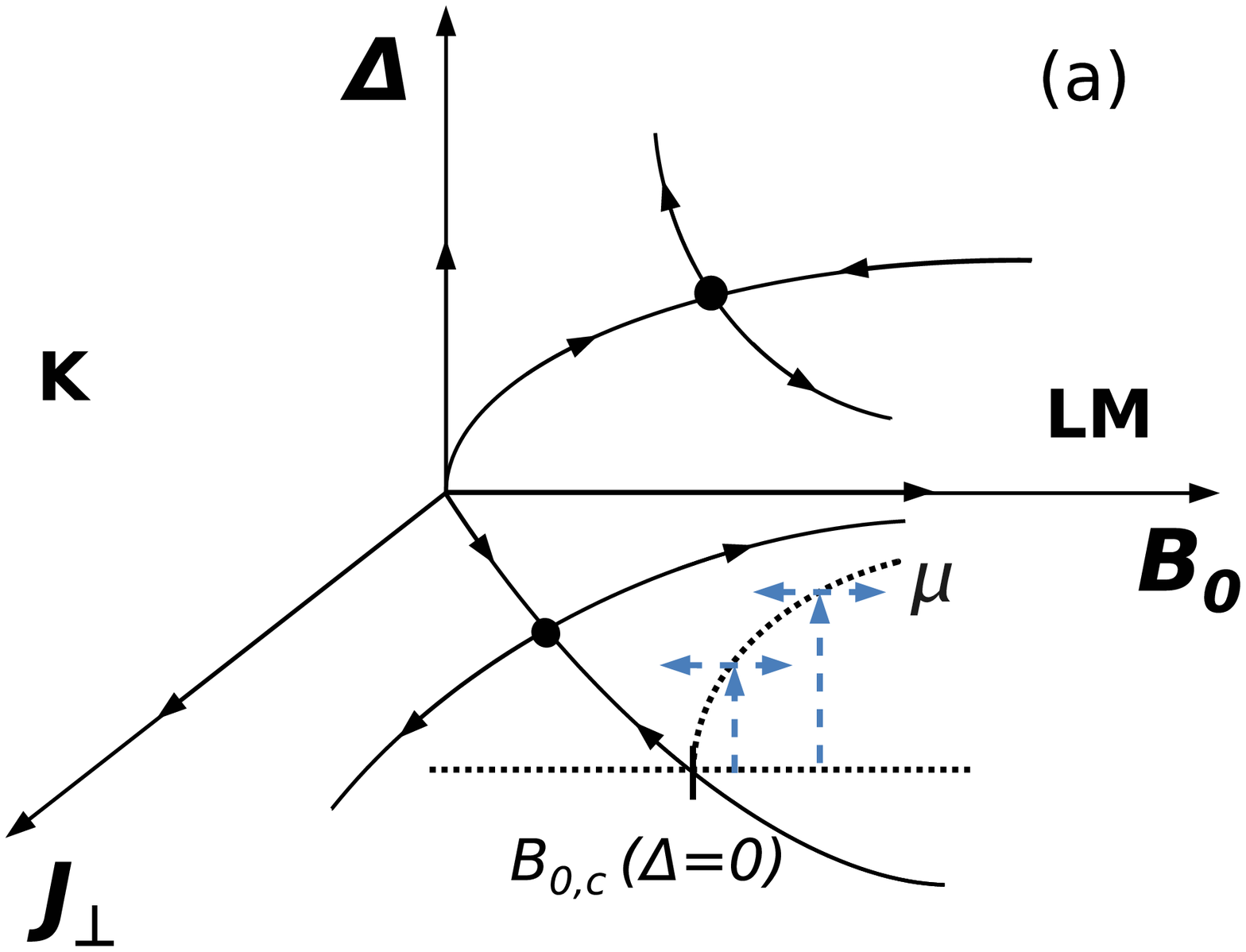}
\includegraphics[width=1.1\columnwidth]{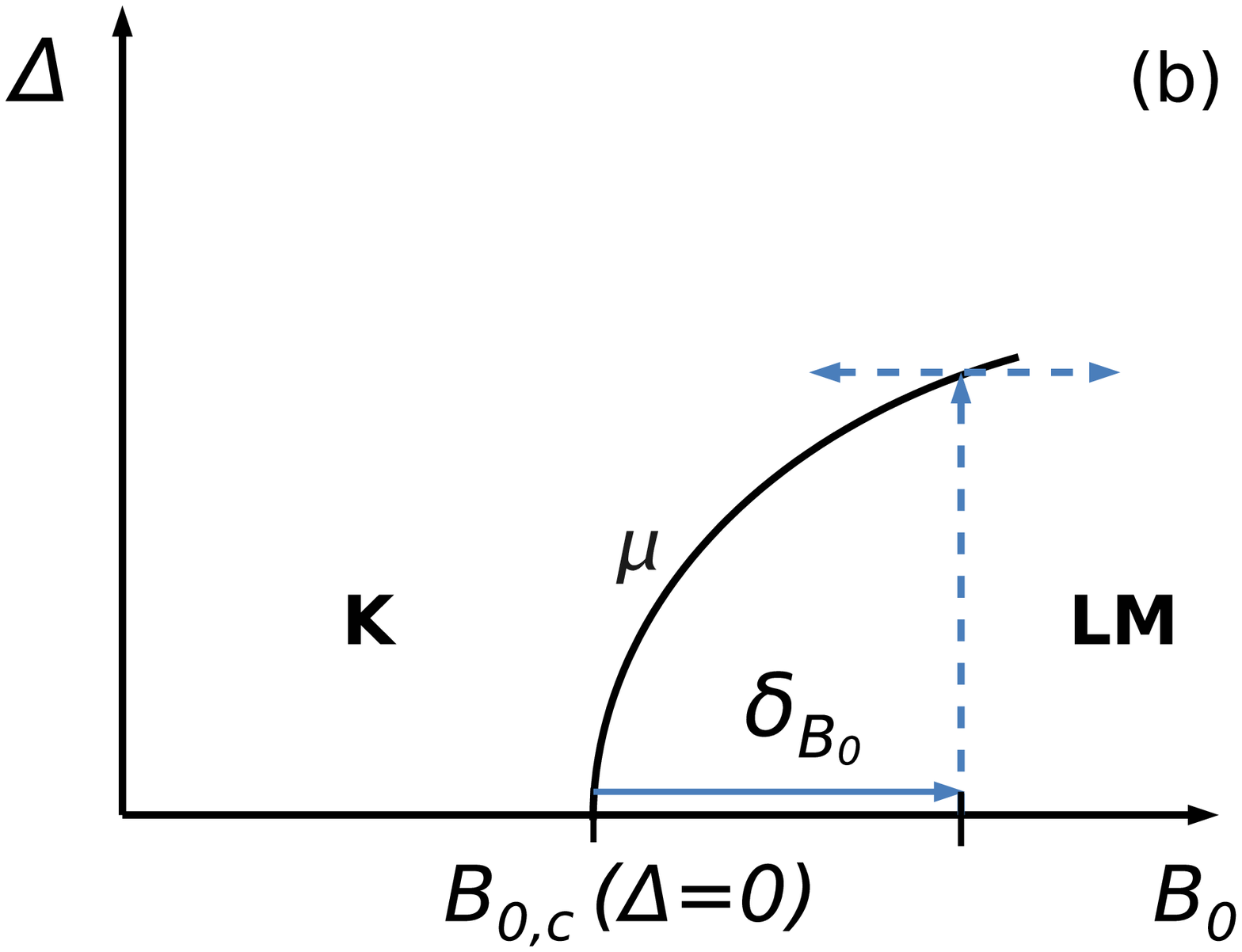}
\includegraphics[width=1.1\columnwidth]{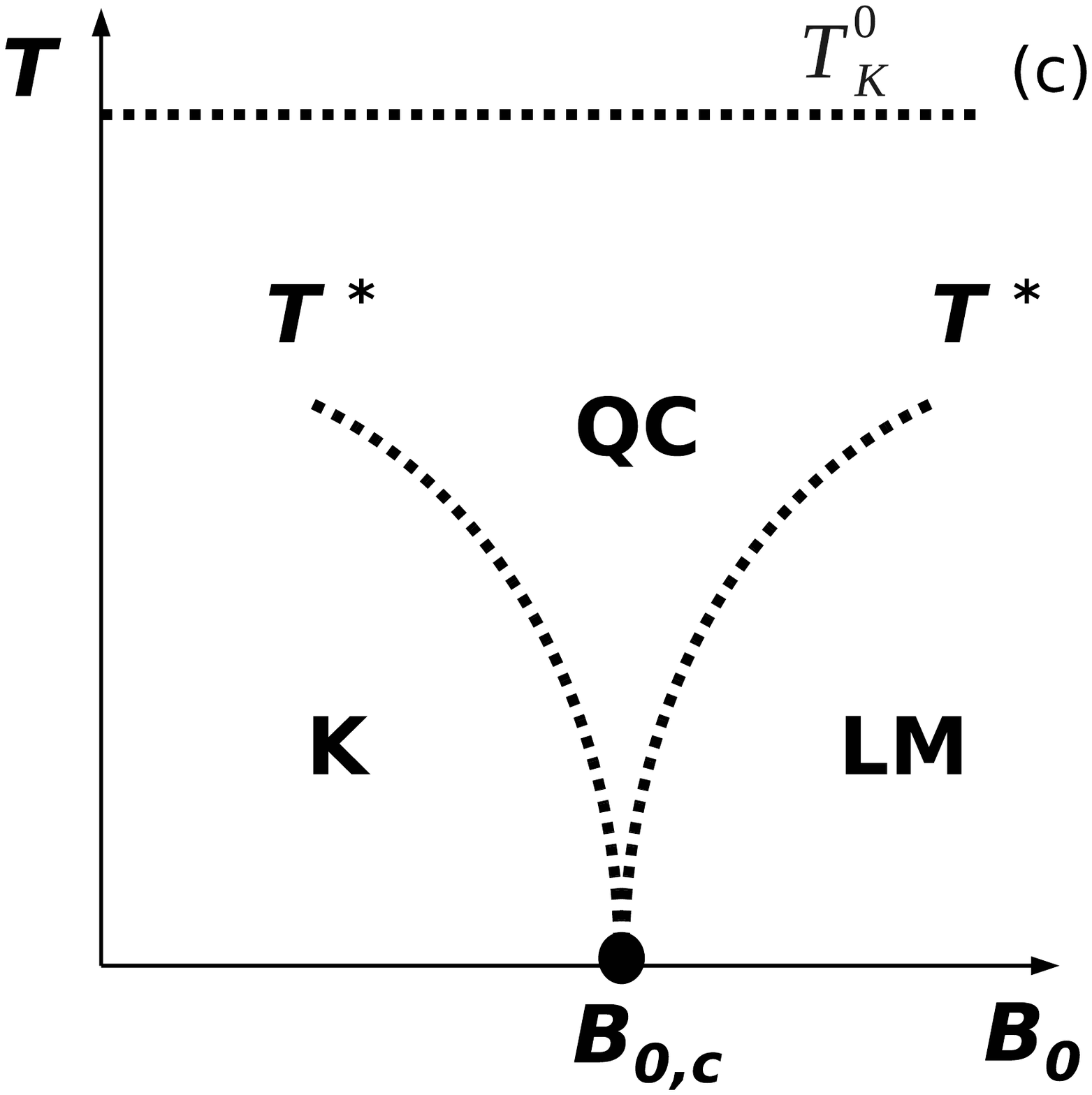}
\caption{(Color online)
(a) Schematic of the projected RG flows for the BFKM and the spin-boson
model in a three-dimensional space spanned by the Kondo spin-flip coupling
$J_{\perp}$, the bosonic coupling $B_0$, and the transverse field $\Delta$. (The
flow of the longitudinal Kondo coupling $J_z$ is not shown.) Arrows on solid
lines show the direction of RG flow and circles represent unstable fixed points.
Arrows on dashed lines indicate the tuning of the bare couplings.
(b) Schematic representation of the procedure of parameter tuning outlined in
the text. 
(c) Schematic $B_0$-$T$ phase diagram for BFKM systems. The upper range of
the quantum-critical (QC) regime is set by $T_K^0$, the Kondo temperature
scale associated with the $B_0=0$ problem.  The boundary between the LM and K phases as shown in (a) and (b) is parameterized by $ \mathbf \mu =( B_{0,c}, \Delta_{c}(B_{0,c}))$ for fixed  $\rho_0 J_K=0.5$.
}
\label{figg2}
\end{figure}

The BFKM with $\Delta=0$ and a sub-Ohmic bath exponent $0<s<1$ features
an unstable fixed point lying on a separatrix in the $B_0$-$J_{\perp}$ plane
[see Fig.\ \ref{figg2}(a)]. 
This critical point governs the transition between
the Kondo (K) phase and the Kondo-destroyed local-moment (LM) phase. 
A transverse field introduces spin flips into the impurity sector, thereby
tending to disfavor the presence of a well-defined local moment. This
leads one to consider whether there will be a quantum phase
transition upon increasing $\Delta$ at fixed $J_K$ and $B_0$.
To address this issue, it is instructive to recall that for $J_K=0$, the
BFKM reduces to the spin-boson model (SBM), where increasing $\Delta$ is known
to drive the system through a second-order quantum phase transition along a
line in the $B_0$-$\Delta$ plane [see Fig.\ \ref{figg2}(a)]. It therefore seems
highly probable for there to be a phase transition in the $B_0$-$\Delta$ plane
at a fixed value $J_{\perp} > 0$, as indicated by the dashed lines in Fig.\
\ref{figg2}(a).
The expected behavior shown in Fig.\ \ref{figg2}(a) also suggests that at a
fixed $\Delta>0$, one should still anticipate encountering a Kondo-destruction
quantum phase transition as $B_0$ is increased.  This, in turn, raises intriguing
questions about the relation between any Kondo-destruction critical points
reached for nonzero $\Delta$ and the $\Delta=0$ BFKM fixed point, and in particular
about the evolution of the critical properties with $\Delta$.

\subsection{Numerical renormalization-group method}

For $\Delta=0$, the BFKM has been treated successfully\cite{Zhu:02} using an
analytical renormalization-group procedure based on an expansion in
$\epsilon=1-s$. It proves to be very difficult to account for a transverse
magnetic field under this approach. By contrast, Eq.\ \eqref{H_BFKM} can be
solved for arbitrary values of $\Delta$ using the Bose-Fermi
extension\cite{Glossop:05+07} of the numerical renormalization
group.\cite{Wilson:75} This section summarizes the Bose-Fermi NRG method and
mentions a few details of its application to the present problem.

The key elements of the method are (i) the division of the continua of band and
bath states into logarithmic bins spanning energy ranges
$\Lambda^{-m+1} < \pm\epsilon/D, \: \omega/\omega_0 < \Lambda^{-m}$ for
$m = 0$, $1$, $2$, $\ldots$, with $\Lambda>1$ being the Wilson discretization
parameter, (ii) an approximation of all states within each bin by a single
representative state, namely, the particular linear combination of states that
couples to the impurity and, (iii) mapping of the problem via the Lanczos method
onto a problem in which the impurity couples only to the end sites of two
nearest-neighbor tight-binding chains, one fermionic and the other bosonic,
having hopping coefficients and on-site energies that decay exponentially along
each chain. These steps yield a Hamiltonian that can be expressed as the limit
$H_{\text{BFKM}}=\lim_{N\to\infty} \alpha \Lambda^{-N/2} D H_N$ of an iterative
sequence of dimensionless, scaled Hamiltonians
\begin{align}
\label{H_N}
H_N
&= \Lambda^{1/2} H_{N-1} \notag \\
&+ \frac{\Lambda^{N/2}}{\alpha} \biggl\{ \sum_{\sigma} \Bigl[ \epsilon_N
   f_{N\sigma}^{\dag} f_{N\sigma}^{\pdag} + \tau_N \bigl( f_{N\sigma}^{\dag}
   f_{N-1,\sigma}^{\pdag} + \text{H.c.} \bigr)\Bigr] \notag \\
&+ \tilde{N} \, \frac{\omega_0}{D} \Bigl[ e_M b_{M}^{\dag} b_M^{\pdag}
   + t_M \bigl( b_M^{\dag} b_{M-1}^{\pdag} + \text{H.c.}
   \bigr)\Bigr] \biggr\} ,
\end{align}
where $\alpha =\frac{1}{2}{\Lambda }^{{1}/{2}}\left(1+{\Lambda }^{-1}\right)$,
$\tilde{N} = N\bmod 2$, and $M=N/2$.
For the flat conduction-band density of states in Eq.\ \eqref{rho:def},
$\epsilon_N = 0$ while $\tau_N\propto\Lambda^{-N/2}$ for $N\gg 1$, meaning that
in Eq.\ \eqref{H_N} the combination $\Lambda^{N/2}\tau_N$ approaches $1$. By
contrast, the bosonic tight-binding coefficients $e_M$ and $t_M$ both vary as
$\Lambda^{-M}$ for $M\gg 1$. In order to treat similar fermionic and bosonic
energy scales at the same stage of the calculation, one site is added to the
end of the fermionic chain at each iteration, whereas the bosonic chain is
extended by the addition of site $M$ only at even-numbered iteration $N=2M$.
The iterative solution begins with the atomic limit described
by $H_0$, where
\begin{align}
\label{H_0}
\alpha H_0
&= \rho_0 J_0 \mathbf{S} \cdot \sum_{\sigma,\sigma'} f_{0\sigma}^{\dag}
   \bsigma_{\sigma,\sigma'} f_{0\sigma'}^{\pdag} + \sum_{\sigma} \epsilon_0
   f_{0\sigma}^{\dag} f_{0\sigma}^{\pdag} \\
&+ \frac{\omega_0}{D}\biggl[\biggl(\frac{B_0}{\pi(s\!+\!1)}\biggr)^{1/2}\!\!S_z
   \bigl( b_0^{\pdag} + b_0^{\dag} \bigr) + e_0 b_0^{\dag} b_0^{\pdag} \biggr]
   + \frac{\Delta}{D} \, S_x. \notag
\end{align}

Many-body eigenstates of iteration $N-1$ are combined with basis states of
fermionic chain site $N$ and (for $N=2M$) bosonic chain site $M$ to form a basis
for iteration $N$. $H_N$ is diagonalized in this basis, and these eigenstates
in turn are used to form a basis for iteration $N+1$. In order to maintain a
basis of manageable dimension, only the $N_s$ eigenstates of lowest energy are
retained after each iteration.\cite{Wilson:75} In Bose-Fermi
problems,\cite{Glossop:05+07} one must also restrict the basis of each bosonic
chain site to the $N_b+1$ number eigenstates $0\le b_M^{\dag} b_N^{\dag} \le N_b$.
For further details, see Refs.\ \onlinecite{Wilson:75} and
\onlinecite{Glossop:05+07}.

In the presence of a transverse magnetic field $\Delta$, no component of the total spin
is conserved. However, $H_N$ commutes with the total ``charge'' operator
\begin{equation}
\label{Q}
\hat{Q} = \sum_{n=0}^N \bigl( f_{n\up}^{\dag} f_{n\up}^{\pdag}
   + f_{n\dn}^{\dag} f_{n\dn}^{\pdag} - 1 \bigr) ,
\end{equation}
Since the conduction-band density of states in Eq.\ \eqref{rho:def} is
particle-hole symmetric, $H_N$ also commutes with
\begin{equation}
\label{isospin}
\hat{I}_+ = \sum_{n=0}^N (-1)^n f_{n\up}^{\dag} f_{n\dn}^{\dag}
\end{equation}
and its adjoint $\hat{I}_- = \hat{I}_+^{\dag}$. In such cases, U(1) charge
conservation symmetry is promoted to an SU(2) isospin symmetry with generators
$\hat{I}_{\pm}$ and $\hat{I}_z=\frac{1}{2} \hat{Q}$.
This symmetry ensures that the many-body eigenstates of $H_N$ can be labeled with
quantum numbers $(I,Q)$ and may be grouped into multiplets of degeneracy $2I+1$.
Moreover, it allows the NRG calculations to be performed using a reduced basis
of states with $Q=-2I$. Our computations took advantage of these symmetry
properties to reduce the labor of obtaining the eigensolution.

Throughout the remainder of this paper, all energies are expressed as multiples
of the half-bandwidth $D=1$. The NRG results presented in the next section were
all obtained for bath exponent $s=0.8$ and dimensionless Kondo coupling
$\rho_0 J = 0.5$. All calculations were performed using a Wilson discretization
parameter $\Lambda=9$, allowing up to $N_b=8$ bosons per site, and retaining up
to $N_s=500$ isospin multiplets after each iteration; all values that were
shown in Ref.
\onlinecite{Glossop:05+07} yield reliable results for the BFKM
without any transverse field.

\section{Kondo-Destruction Quantum Phase Transition}
\label{Sec:QPT}

In this section, we determine the phase diagram of the Ising-anisotropic BFKM
in the presence of a local transverse field. We focus attention on the
Kondo-destruction quantum phase transition, which separates a Kondo (K) phase
and a Kondo-destroyed local-moment (LM) phase.

\subsection{Critical Kondo Destruction}

We began by locating the quantum phase transition of the BFKM in the absence
of a transverse field, expressed as a critical coupling to the
bosonic bath $B_{0,c}(\Delta\!=\!0)$ (for the fixed value $\rho_0 J_K=0.5$ used
throughout this work). The signature of proximity to a critical point is the flow
of the NRG many-body eigenenergies over some finite range of intermediate iterations
$N$ (corresponding to a window of temperatures $T\sim\Lambda^{-N/2}$) to new values
distinct from the converged eigenenergies obtained for large $N$. The latter spectra
describe the stable RG fixed points of the problem, which in this
case govern the K and LM phases. As $B_0$ is tuned closer to
$B_{0,c}(\Delta\!=\!0)$, the energies remain flat and close to the new values over
an increasingly wide range of $N$.

Once $B_{0,c}(\Delta\!=\!0)$ was determined, we considered a sequence of five
larger bosonic couplings 
\begin{equation}\label{eq42}
B_0 = B_{0,c}(\Delta\!=\!0) + \delta_{B_0} ,
\end{equation}
where $\delta_{B_0}=6\times 10^{-n}$ with $n = 1, \ldots, 5$. For each value of $\delta_{B_0}$, we increased the transverse field from zero until
we reached a new critical point, again judged by examination of the many-body
spectrum. Once the critical field $\Delta_c(\delta_{B_0})$ was established, we
fixed $\Delta=\Delta_c$ and tuned $B_0$ around the value
$B_{0,c}(\Delta_c)\equiv B_{0,c}(0) + \delta_{B_0}$
in order to calculate the critical properties
associated with the finite-$\Delta$ transition.
The procedure is illustrated schematically in Fig.\ \ref{figg2}(b).

\begin{figure}[!h]
\includegraphics[width=1.1\columnwidth]{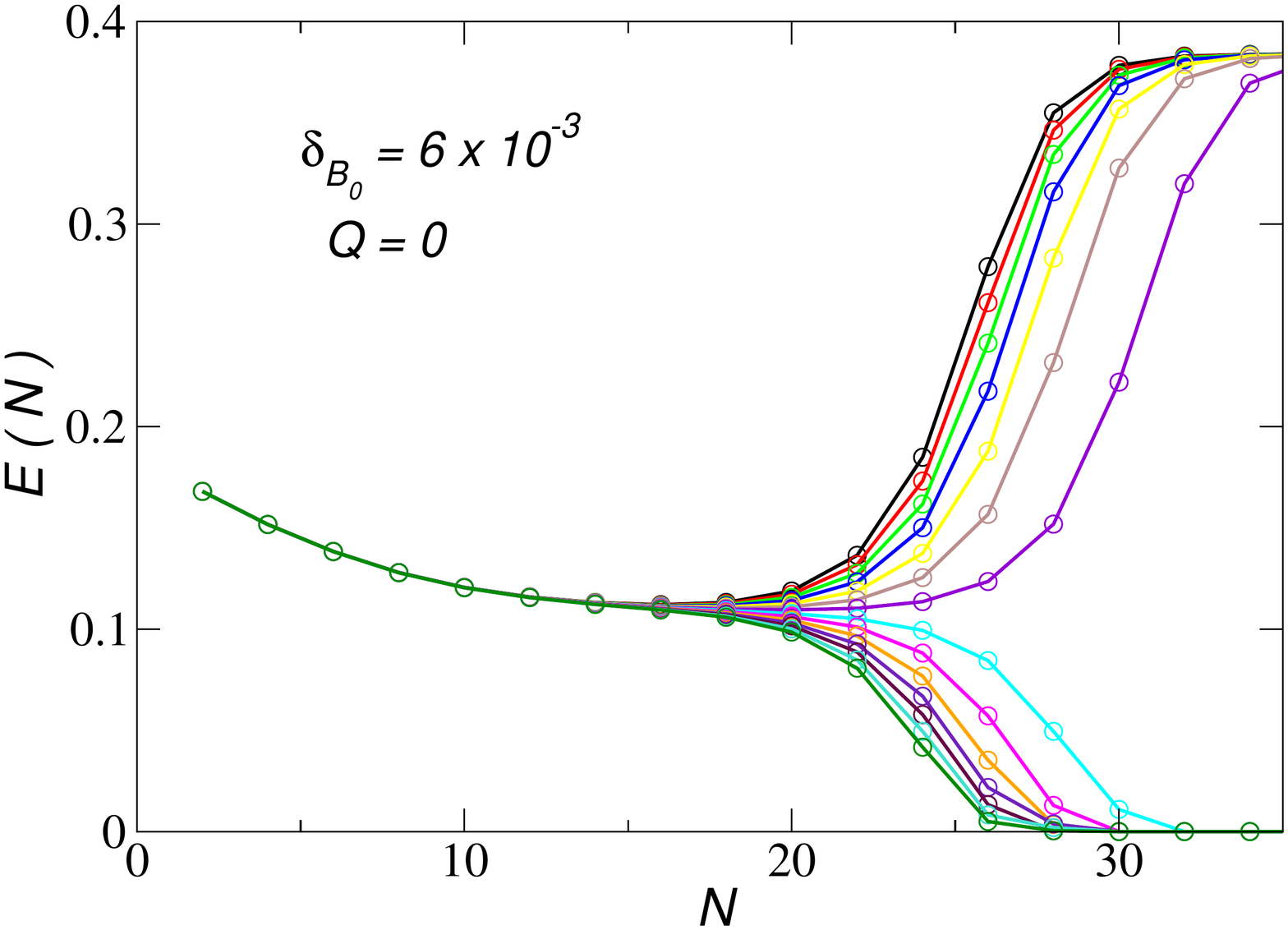}
\caption{The lowest bosonic ($Q=0$) eigenenergy as a function of even iteration
number for $\delta_{B_0}=6\times{10}^{-3}$. The energies converging to
$E \ \simeq 0.39$ correspond to effective bosonic couplings
$-9.45\times 10^{-6}\le B_0-B_{0,c}\le -8.5\times 10^{-7}$, while the energies
converging to $E=0$ span the range
$8.5\times 10^{-7}\le B_0-B_{0,c} \le 9.45\times 10^{-6}$.
}
\label{fig3}
\end{figure}

Since all values of $\delta_{B_0}$ show qualitatively the same behavior, we
concentrate below on the representative case ${\delta }_{B_0}=6\times{10}^{-3}$.
The charge $Q$ defined in Eq.\ (\ref{Q}) is a good quantum number, so it can be used
to label the different NRG many-body eigenstates. To begin with, we consider states having $Q=0$. (The $Q=1$ states are discussed
 in detail in Sec.\ IV.)
Figure \ \ref{fig3} shows the energy of the lowest excited $Q=0$ state as a function
of iteration number for $\delta_{B_0}=6\times{10}^{-3}$. The spectrum is
different for odd $N$, not only due to the alternation properties of fermions on finite
tight-binding chains, but also  because bosons are added only at even 
iterations.\cite{Glossop:05+07} Since the many-body ground state shares the
quantum number $Q=0$, we provisionally interpret the lowest excited state of
the same charge as having bosonic character.

From Fig. 3, one can clearly distinguish the quantum-critical regime at
intermediate values of $N$ and the crossover for large $N$ to either
the K or the fixed LM point. For $B_0 < B_{0,c}$, the first bosonic
excitation energy eventually approaches the value $E\simeq 0.39$ that
it takes in the free-boson spectrum obtained for $B_0=0$. For
$B_0 > B_{0,c}$, the first bosonic excitation energy eventually
vanishes, reflecting the two-fold degeneracy of the ground state in the
LM phase. In the quantum-critical regime, the excitation energy takes
a distinct value $E\simeq 0.11$ that can be considered a characteristic
of the QCP.

The crossover to the spectrum of either the LM or the K fixed point can be
used to estimate the crossover scale $T^*\sim\Lambda^{-N^*/2}$, where $N^*$ is
the iteration at which the difference of the eigenvalues from their critical
values exceeds a predetermined threshold.
The lowest bosonic eigenvalue in Fig.\ \ref{fig3} shows that $T^*$ vanishes at the
critical point from both sides since $N^*\to\infty$ as $B_0\to B^{\pm}_{0,c}$.
The schematic $B_0$-$T$ phase diagram is shown in Fig.\ \ref{figg2}(c).

\begin{figure}[!h]
\includegraphics[width=1.1\columnwidth]{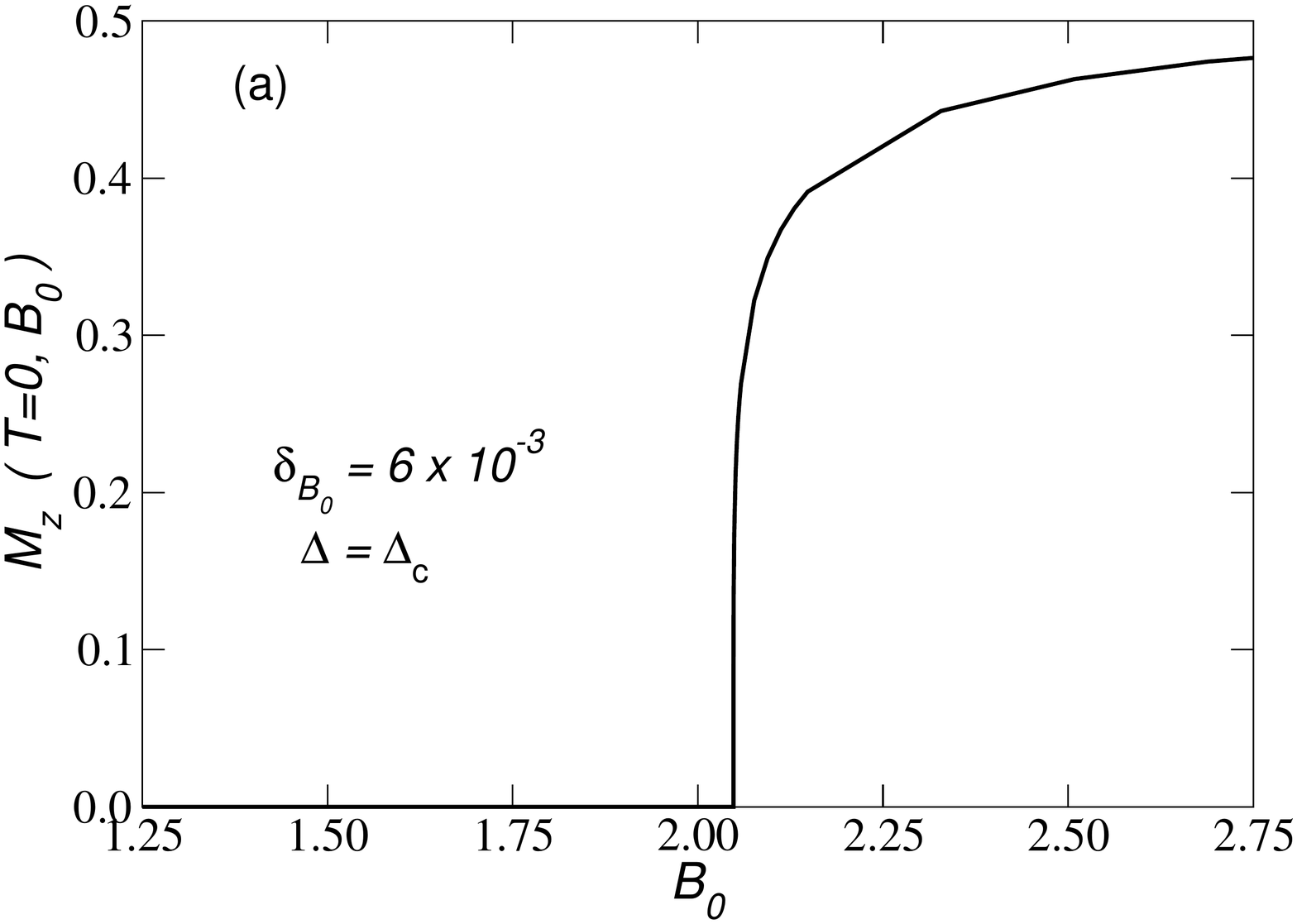}
\includegraphics[width=1.1\columnwidth]{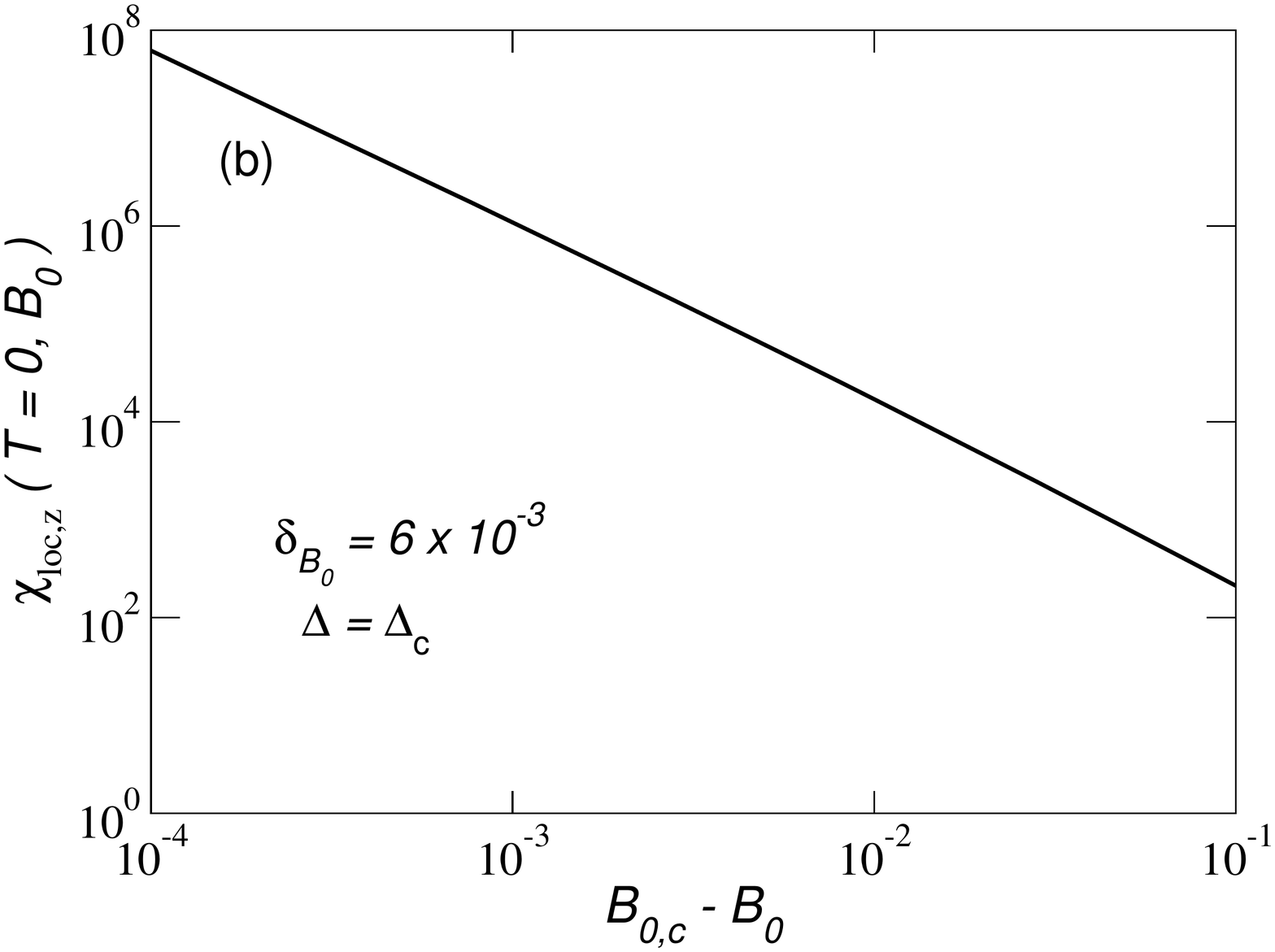}
\includegraphics[width=1.1\columnwidth]{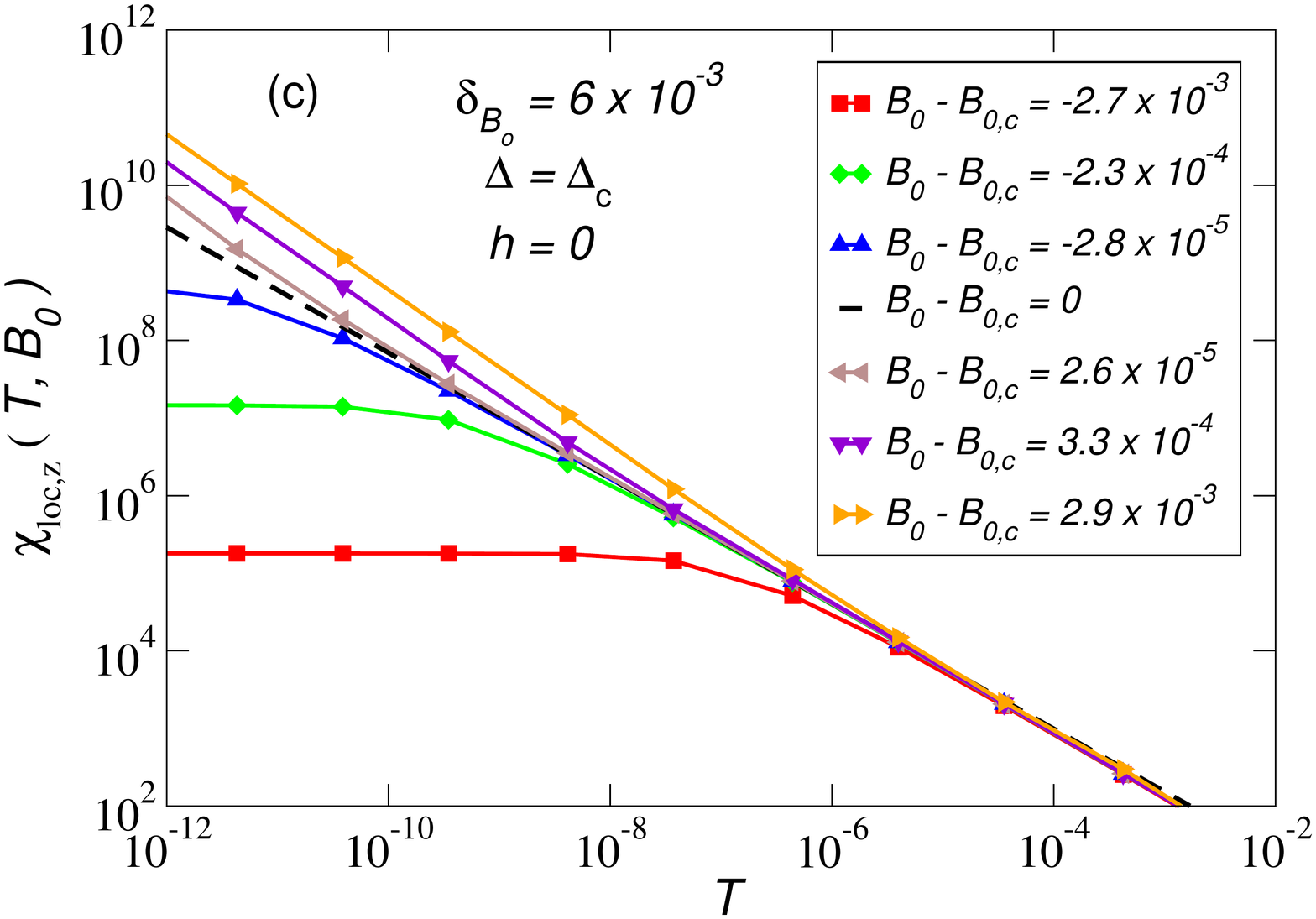}
\caption{(a) Local magnetization $M_z$ vs $B_0$ across the quantum phase
        transition. (b) Zero-temperature longitudinal susceptibility
        $\chi_{\mathrm{loc},z}$ in the Kondo phase near the critical point.
        (c) $\chi_{\mathrm{loc},z}$ vs $T$ for $B_0$ values at (dashed line)
        and near (symbols) $B_{0,c}$. All data are for
        $\delta_{B_0}=6\times 10^{-3}$.}
\label{fig4}
\end{figure}

The local magnetization, defined as 
\begin{equation}\label{eq43}
M_z=\mathop{\lim}_{h\to 0,\ T\to 0} \left\langle S_{z}\right\rangle
\end{equation}
where $h$ is a longitudinal magnetic field, is nonzero in the LM phase, vanishes continuously as
$B_0\to B^+_{0,c}$, and remains zero throughout the K phase, as
shown in Fig.\ \ref{fig4}(a). 

\subsection{Critical Exponents}

The local susceptibility in the $z$-direction defined as 

\begin{equation}\label{eq44} 
\chi_{loc,z}\left(T;\omega =0\right)=-{\left.\frac{\partial \left\langle S_z\right\rangle }{\partial h}\right|}_{h=0}={\mathop{\lim }_{h\to 0} \left(-\frac{\left\langle S_z\right\rangle }{h}\right)\ }
\end{equation}
\noindent diverges at $T=0$, $B_0\to B^-_{0,c}$ as shown in Fig.~\ref{fig4}(b).

The distinctions between the LM, K and QC regimes 
[see Fig.~\ref{figg2}(c)]
can also be seen by analyzing 
$\chi_{loc,z}(B_0,T)$ 
in the corresponding regimes. Figure~\ref{fig4}(c) shows 
$\chi_{loc,z}(B_0,T)$ 
for values of $B_0$ below, above and close to $B_{0,c}$. For $T>T^*$, flows from both $B_0<B_{0,c}$ and $B_0>B_{0,c}$ behave the same way by scaling as $T^{-x}$, while for $T<T^*$,  the K fixed point is characterized by a constant Pauli susceptibility and the LM fixed point susceptibility follows a $1/T$  Curie-Weiss form.

\begin{figure}[!h]
\includegraphics[width=1.1\columnwidth]{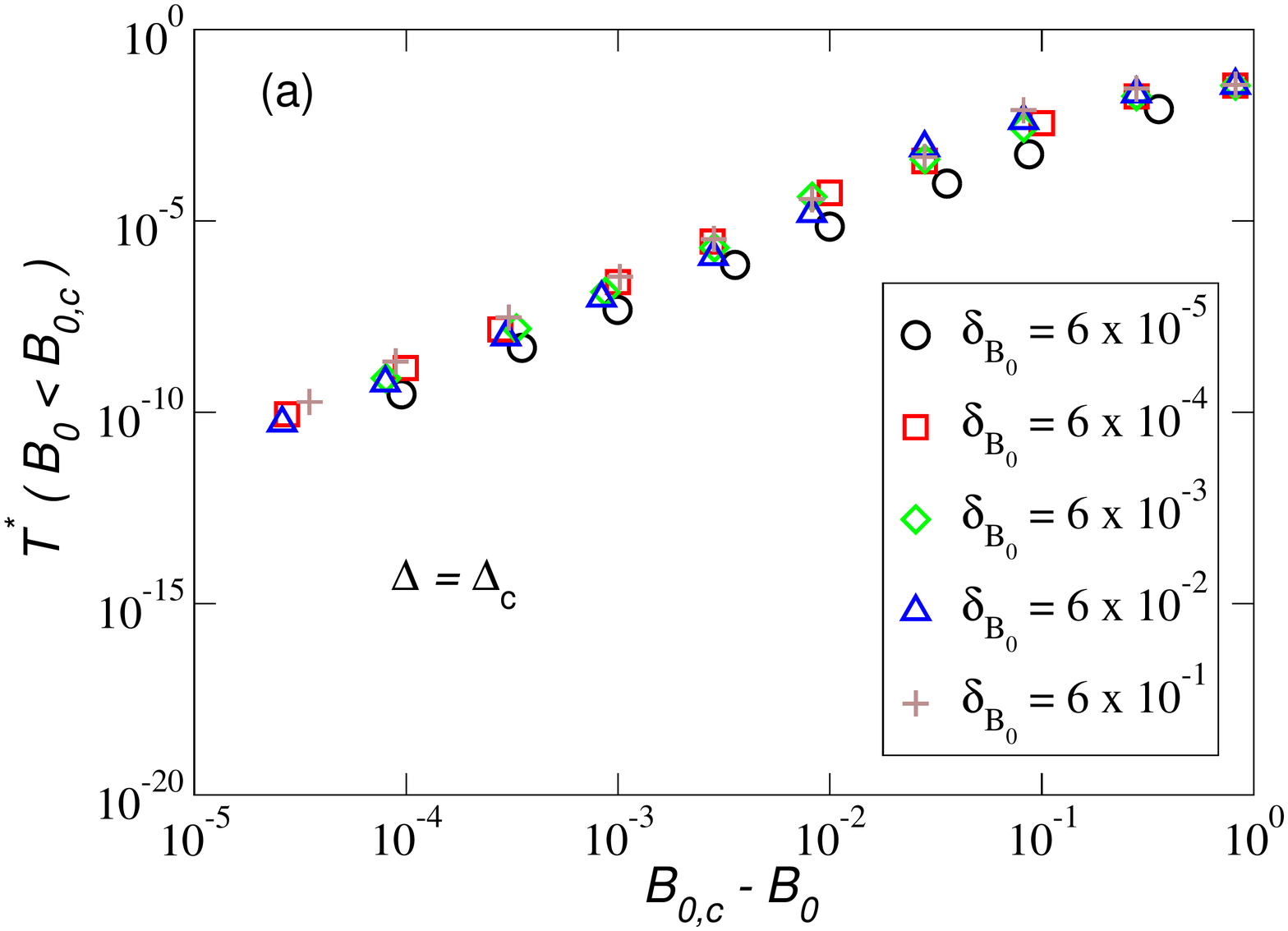}
\includegraphics[width=1.1\columnwidth]{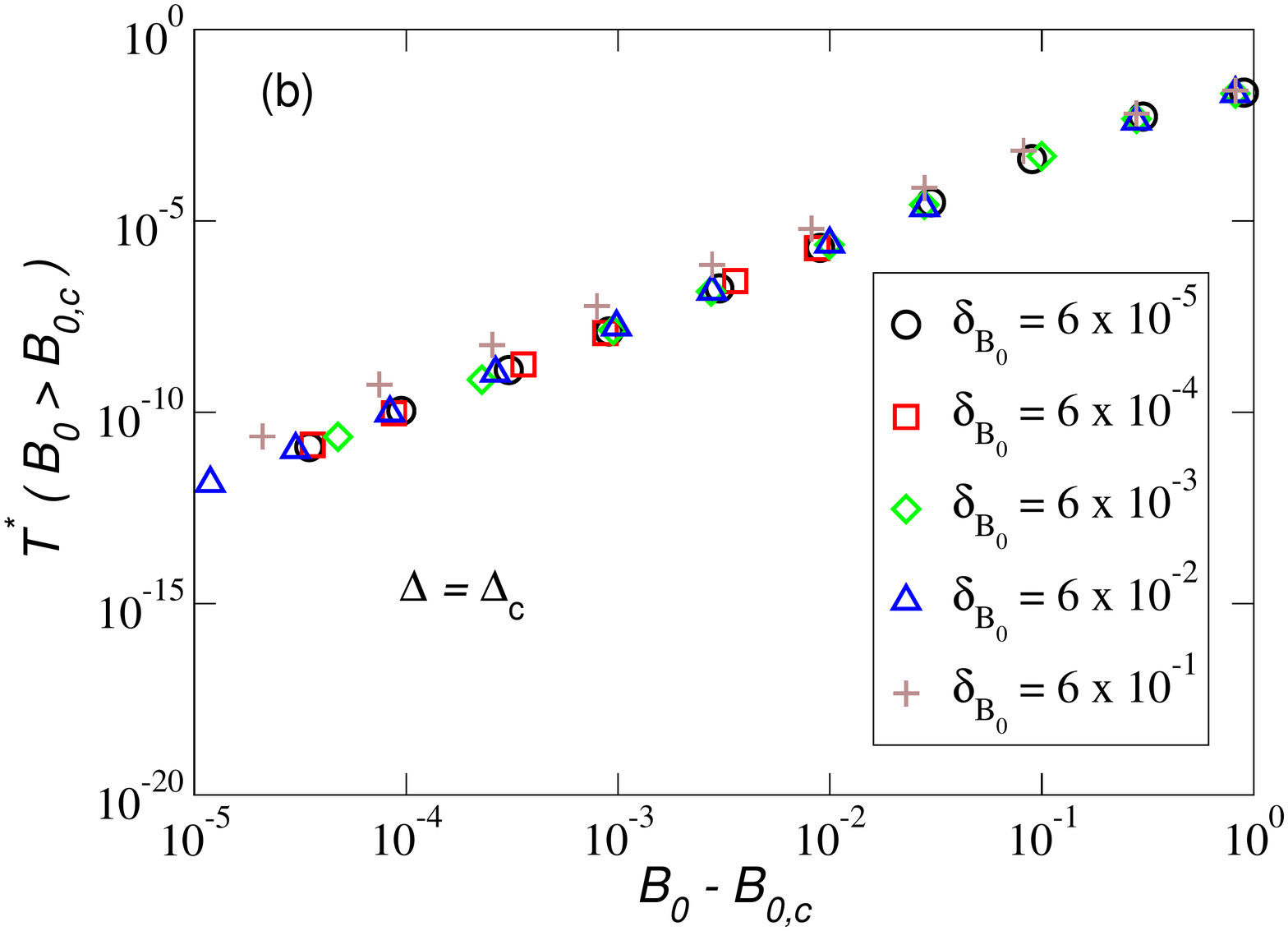}
\caption{ Scaling of $T^*$ with $B_0$ on the K side (a) and on the LM side (b) of the QCP.}
\label{fig5}
\end{figure}

 We have extracted critical exponents for all five investigated values.
 The crossover scale $T^*$ vanishes from both the LM and K phases as

\begin{equation}\label{eq45}
T^*\propto {\left|B_0-B_{0,c}\right|}^{\nu z}
\end{equation}

\noindent with $z=1$ for an impurity problem. 
$T^*$ was estimated from the crossover iteration number $N^*$ as explained in Sec. III.A.
 Figures~\ref{fig5}(a) and~\ref{fig5}(b) show the scaling of  $T^*$ with distance from $B_{0,c}\ $from the K and LM sides, respectively. It is apparent that for all values of ${\delta }_{B_0}$ considered, $\nu $ is essentially the same.

\begin{figure}[!h]
\includegraphics[width=1.1\columnwidth]{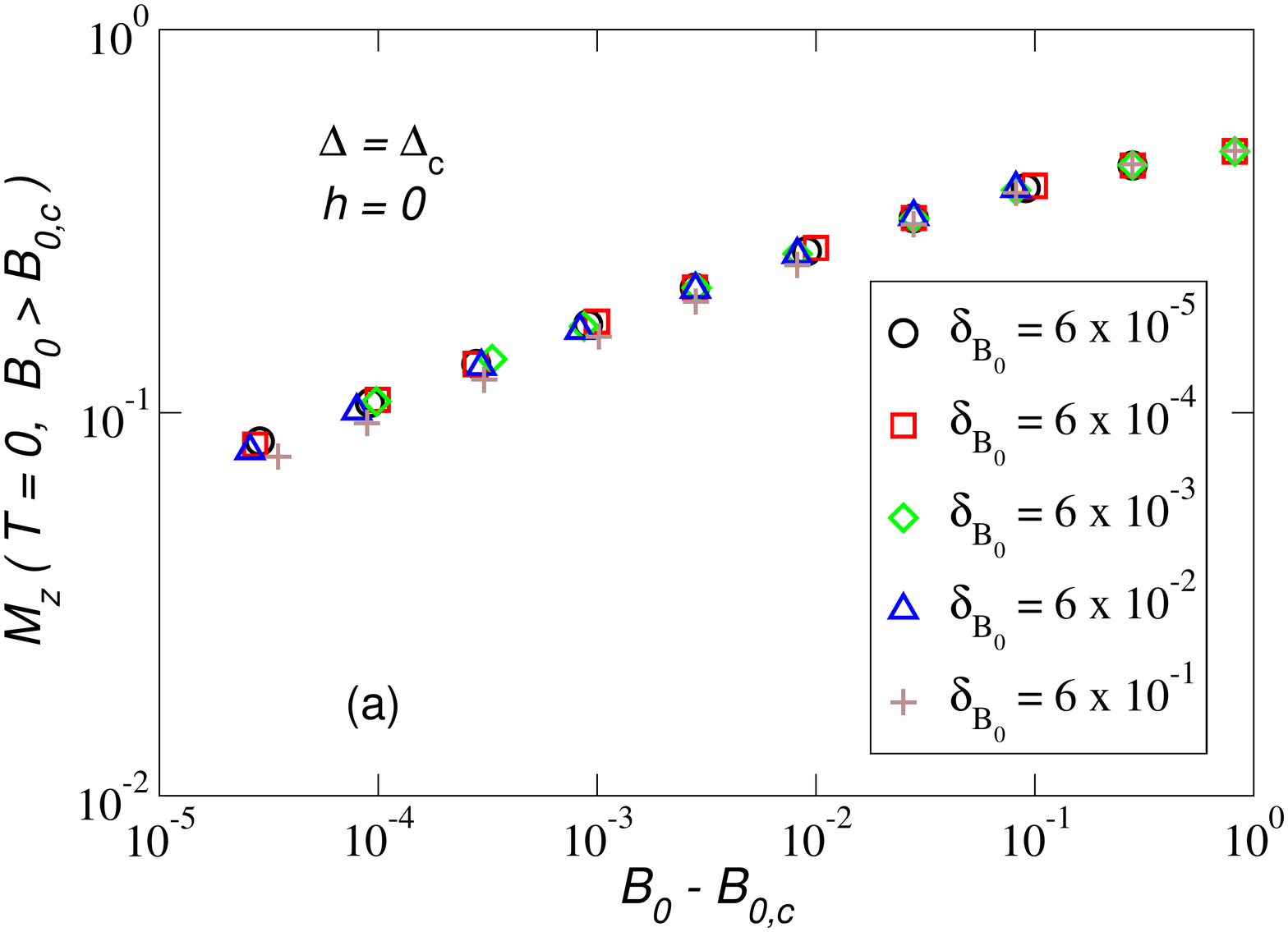}
\includegraphics[width=1.1\columnwidth]{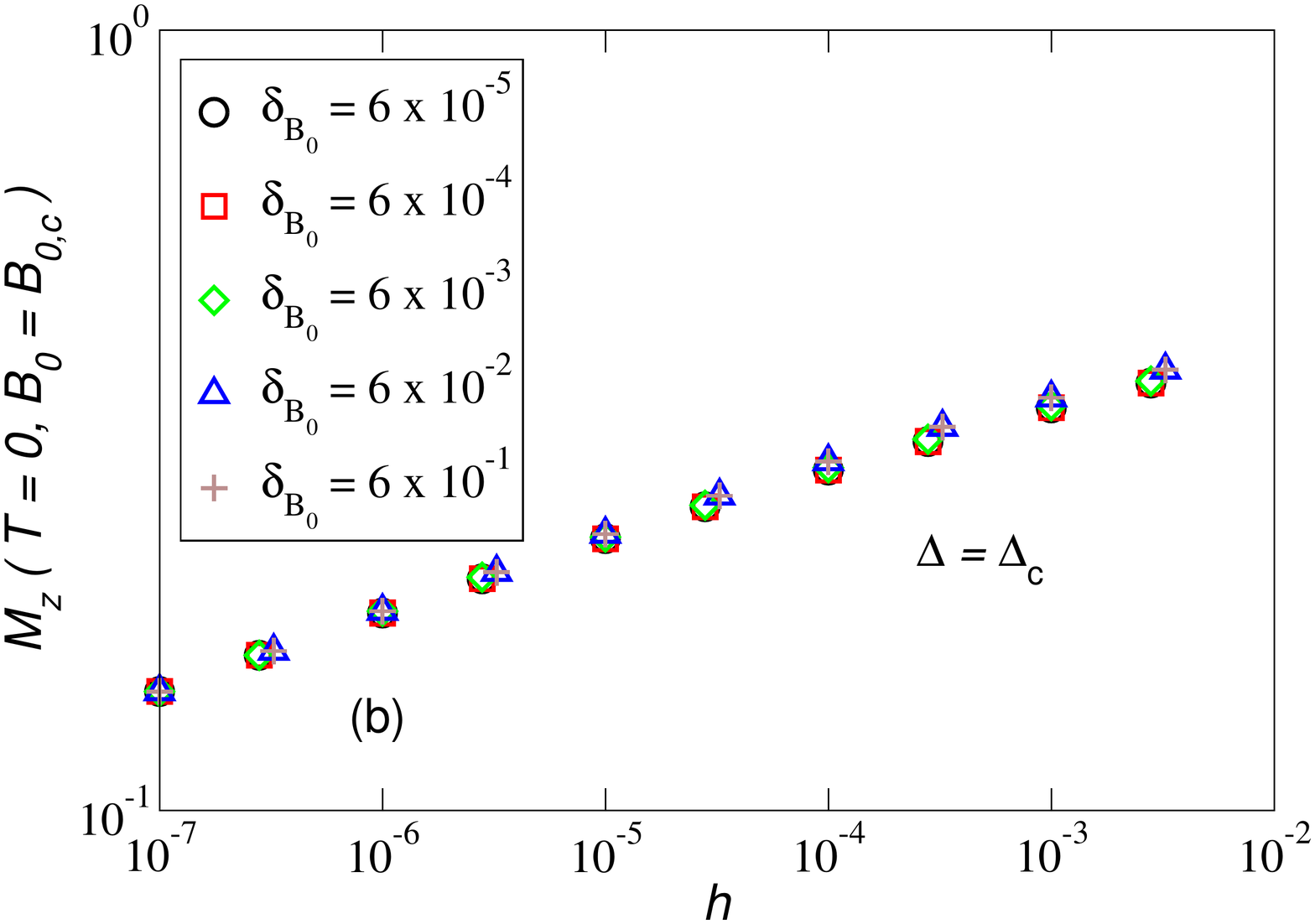}
\caption{(a) Scaling of $M_{z}$ with $B_0$ from LM side. (b) Scaling of $M$ with $h$ at $B_0=B_{0,c}$. 
}
\label{fig6}
\end{figure}

  It is found that $M_z(T=0)$ decays as a power law on
the LM side:
\begin{equation}\label{eq46}
M_z(h\to 0, T\to 0,\ \Delta \to {\Delta }_c)\propto {\left(B_0-B_{0,c}\right)}^{\beta },
\end{equation}

\noindent where $h\ $is an infinitesimal field applied along the $z$ direction as can be seen in Fig.~\ref{fig6}(a)  for all ${\delta }_{B_0}$. It also scales with $h$ at $B_0\simeq B_{0,c}$:

\begin{equation}\label{eq47}
M_z(h,\ B_0=B_{0,c},T=0)\ \propto {\left|h\right|}^{1/\delta}.
\end{equation}
\noindent This is shown in Fig.~\ref{fig6}(b) for all ${\delta }_{B_0}$increments.

\begin{figure}[!h]
\includegraphics[width=1.1\columnwidth]{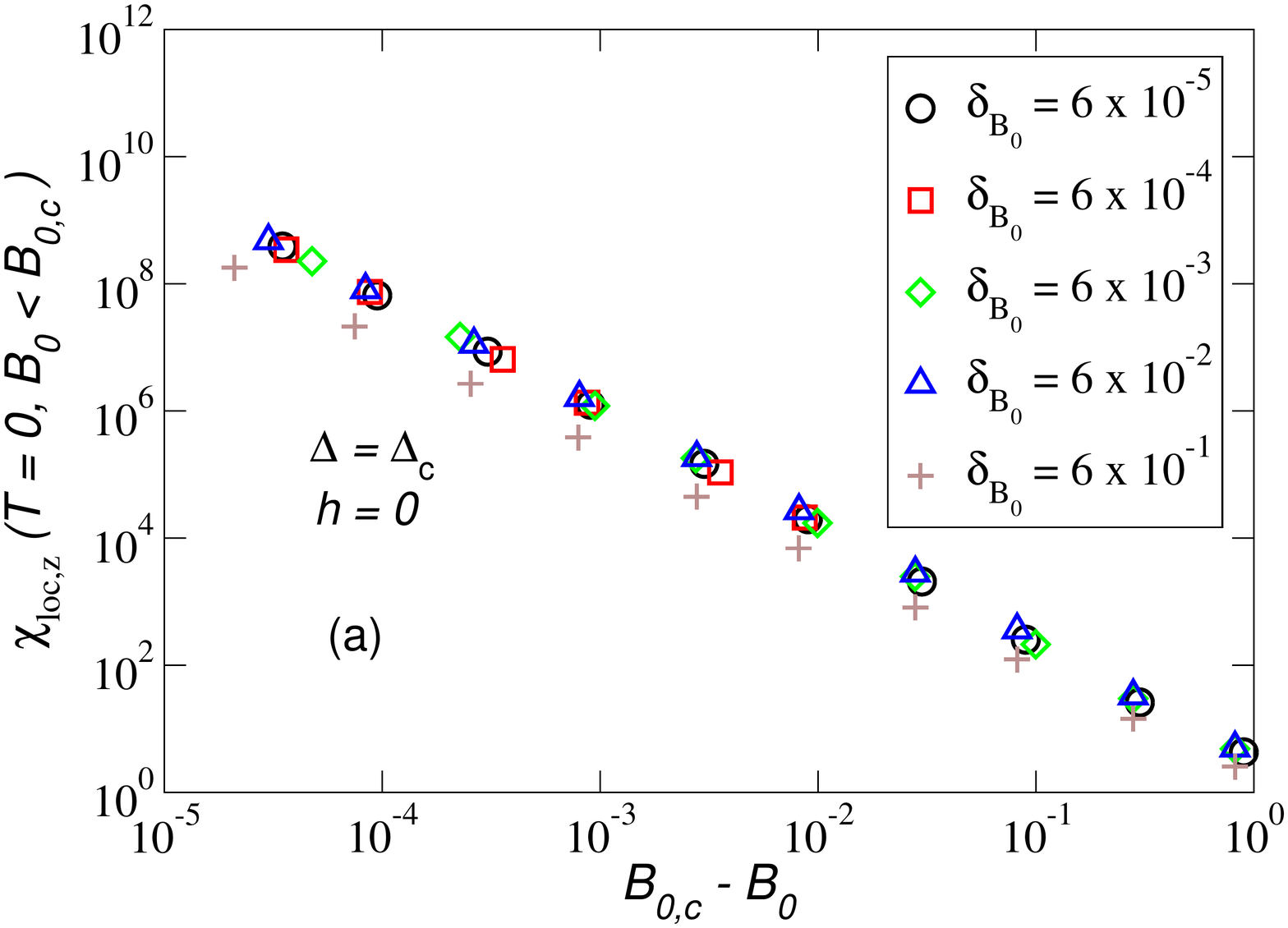}
\includegraphics[width=1.1\columnwidth]{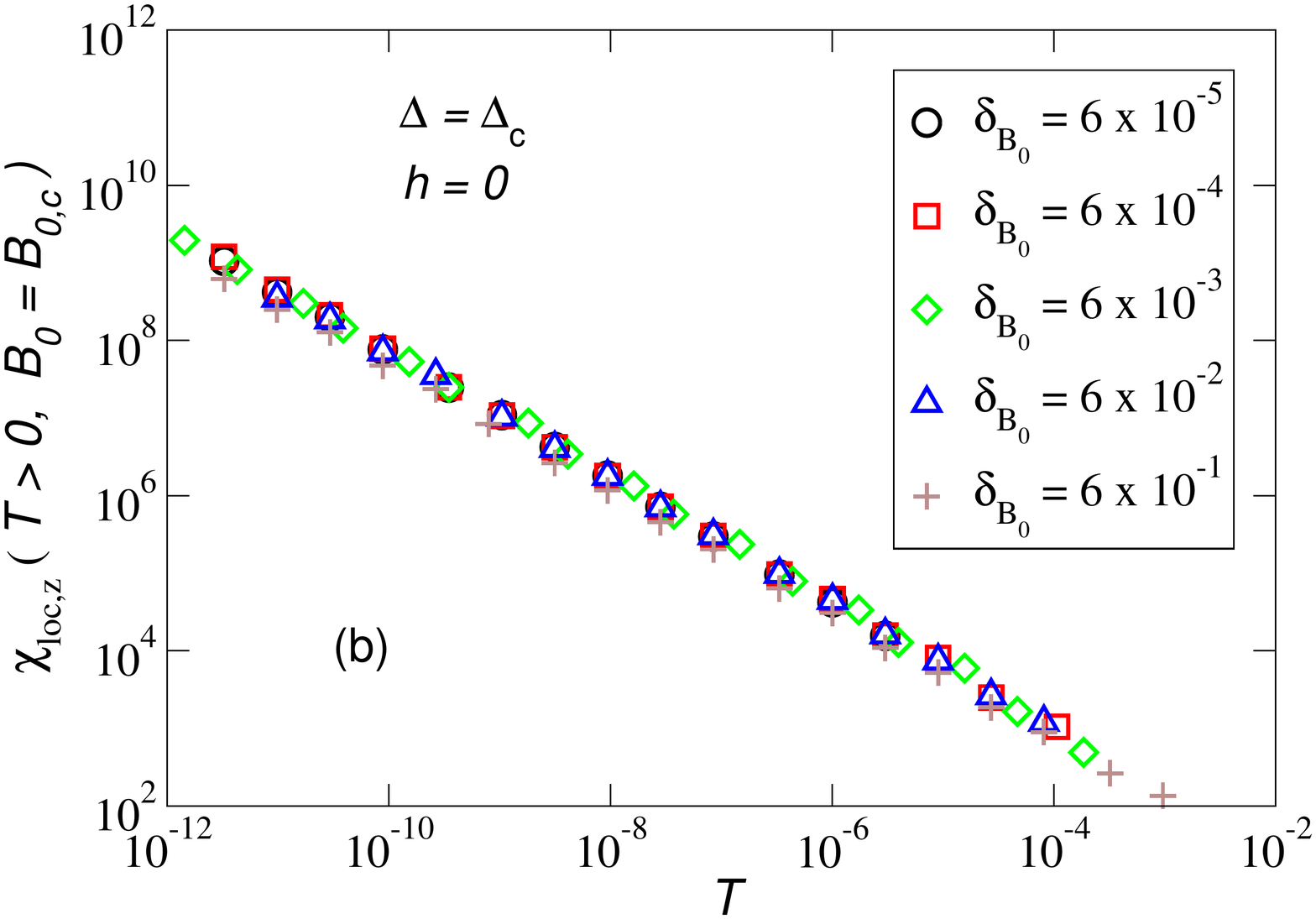}
\caption{(a) Scaling of ${{\rm \chi}}_{loc,z}$ with $B_0$ 
on the K side of the QCP. (b) Scaling of $\chi_{loc,z}$ with $T$ at $B_0=B_{0,c}$.
}
\label{fig7}
\end{figure}

 The local susceptibility in the $z$ direction diverges on approach to the QCP 
from the K side at $T=0$:

\begin{equation}\label{eq48}
{{\rm \chi}}_{loc,z}(B_0<B_{0,c},T=0)\propto {\left(B_{0,c}-B_0\right)}^{-\gamma }
\end{equation}

\noindent and scales with temperature as 
\begin{equation}\label{eq49}
{{\rm \chi}}_{loc,z}(B_0=B_{0,c},T>T^*\ )\propto T^{-x},
\end{equation} 
\noindent as illustrated in Figs.~\ref{fig7}(a) and \ref{fig7}(b) respectively.

\begin{figure}[!h]
\includegraphics[width=1.1\columnwidth]{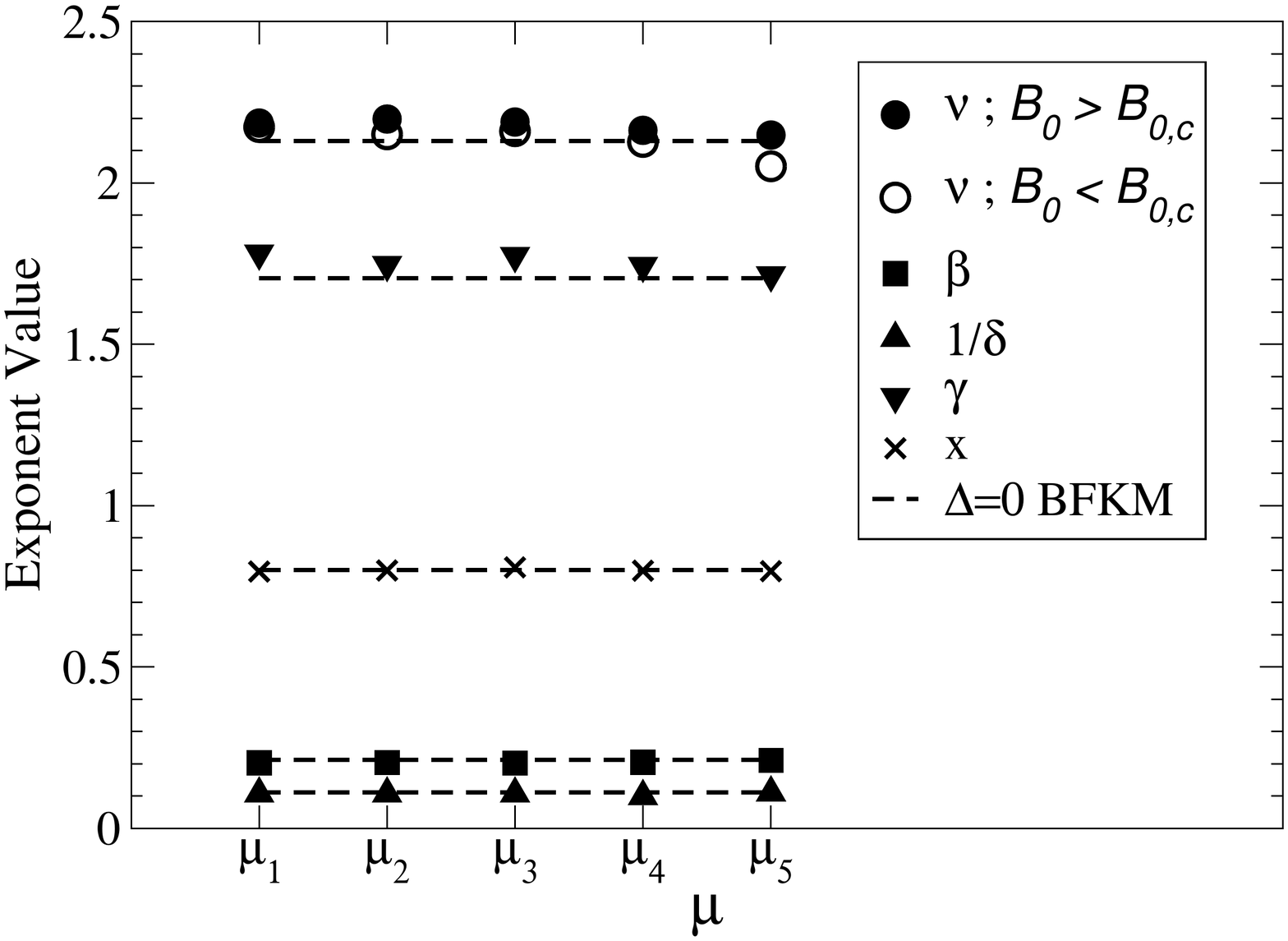}
\caption{Calculated critical exponents. The exponents at the critical coupling $\Delta_c$ corresponding to different
choices of $\delta_{B_0}$ and for fixed $\rho_0 J_K=0.5$ as illustrated in  Figs.~\ref{figg2}(a) and 2(b). The ordinate $\mu_1$ corresponds to $ \delta_{B_0} = 6 \times 10^{-5} $,  $ \mu_2 $ to 
$ \delta_{B_0} = 6 \times 10^{-4} $, and so on. The dashed lines show the corresponding exponents for the $\Delta=0$ BFKM, with $x$ and $\beta$ being NRG results and the remaining exponents derived assuming hyperscaling.}
\label{fig8}
\end{figure}

The values of
all the calculated critical exponents for all values of ${\delta }_{B_0}$ are shown in Fig.~\ref{fig8}. The parameter $ \mathbf \mu =( B_{0,c}, \Delta_{c}(B_{0,c}))$ defines the boundary between the LM and K phases for fixed  $\rho_0 J_K=0.5$ as shown in Figs.~\ref{figg2}(a) and~\ref{figg2}(b).

We note that the critical exponents are virtually the same as those for the $\Delta=0$ BFKM, indicated using dashed lines in Fig. 8. It was previously found\cite{Glossop:05+07} that the 
exponents obey the hyperscaling 
relations derived from the ansatz

\begin{equation}\label{eq50}
F_{crit}=Tf\left(\frac{\left|\Delta -{\Delta }_c\right|}{T^{1/\nu}},\ \frac{\left|h\right|}{T^b}\right),
\end{equation}
\noindent namely 

\begin{equation}\label{eq51}
\delta =\frac{1+x}{1-x},
\end{equation}
\begin{equation}\label{eq52}
\beta =\frac{1}{2}\nu \left(1-x\right),
\end{equation}
\begin{equation}\label{eq53}
\gamma =\nu x
\end{equation}

\noindent  Since the $\Delta>0$ critical exponents coincide for all the values of $\delta_{B_0}$ with their $\Delta=0$ counterparts, hyperscaling also holds in the presence of a transverse field.

We also stress that these conclusions are restricted to the case of bath exponent $s=0.8$ treated in this
paper. Studies for other values of $s$ will be reported elsewhere.

 The error estimates and other related issues are discussed in the Appendix.
 
 \section{Line of Kondo-destruction fixed points} \label{Sec:LFP}

Having established the existence of Kondo-destruction transitions in the presence of a transverse local
magnetic field $\Delta$, we turn next to the relationship among the critical points for different values of $\Delta$.

One possibility is that the RG flows are away from the $\Delta {\rm =0\ }$BFKM\textbf{ } critical fixed point towards a different unstable fixed point, which effectively governs the transitions for all the ${\delta }_{B_0}$considered. 

A plausible conjecture is that these flows take the system from the critical fixed point of the $\Delta=0$ BFKM to the critical
fixed point of the SBM
[See Fig.~\ref{figg2}(a)]. 
The true 
asymptotic critical behavior will be that associated with the SBM model, where the transverse field is solely responsible for quantum-mechanical tunneling with a critical suppression of Kondo tunneling (${\rho J}^*_{\bot }=0)$. Other related scenarios are of course possible {\it a priori.}

A second possibility is that there is a line of unstable fixed points extending from the original $\Delta {\rm =0\ }$BFKM. In this case, we expect that each of the trial ${\delta }_{B_0}$\textbf{ }
is tuned to its own unstable fixed point. We believe this is the correct picture. 

An indication in favor of the second possibility comes from the critical many-body NRG spectrum extracted from the flows on the verge of crossing over to either stable fixed point. The many-body spectrum can be decomposed into a superposition of distinct fermionic and bosonic excitations at any of the two stable fixed points.\cite{Glossop:05+07} Here, the lowest $Q=1$ excitations correspond to single-particle charged fermionic excitations, while the lowest $Q=0$ excitations are single
 boson excitations. Note also that in both of the above cases, one can have higher energy charge $0$ particle-hole excitations. The excitation energy can be written as $E_{Exc}=E_{Fermion}+E_{Boson}$. At the LM fixed point $E_{Boson}$ consists of a superposition of free-boson excitations, while $E_{Fermion}$ is an energy from the spectrum of fermions undergoing spin-dependent potential scattering. At the K fixed point, $E_{Boson}$ is given by free-boson excitations and $E_{Fermion}$ is 
the energy of one or more Kondo-like quasiparticles. Such a sharp distinction cannot be established rigorously from the flows close to the critical point.

\begin{figure}[!h]
\includegraphics[width=1.1\columnwidth]{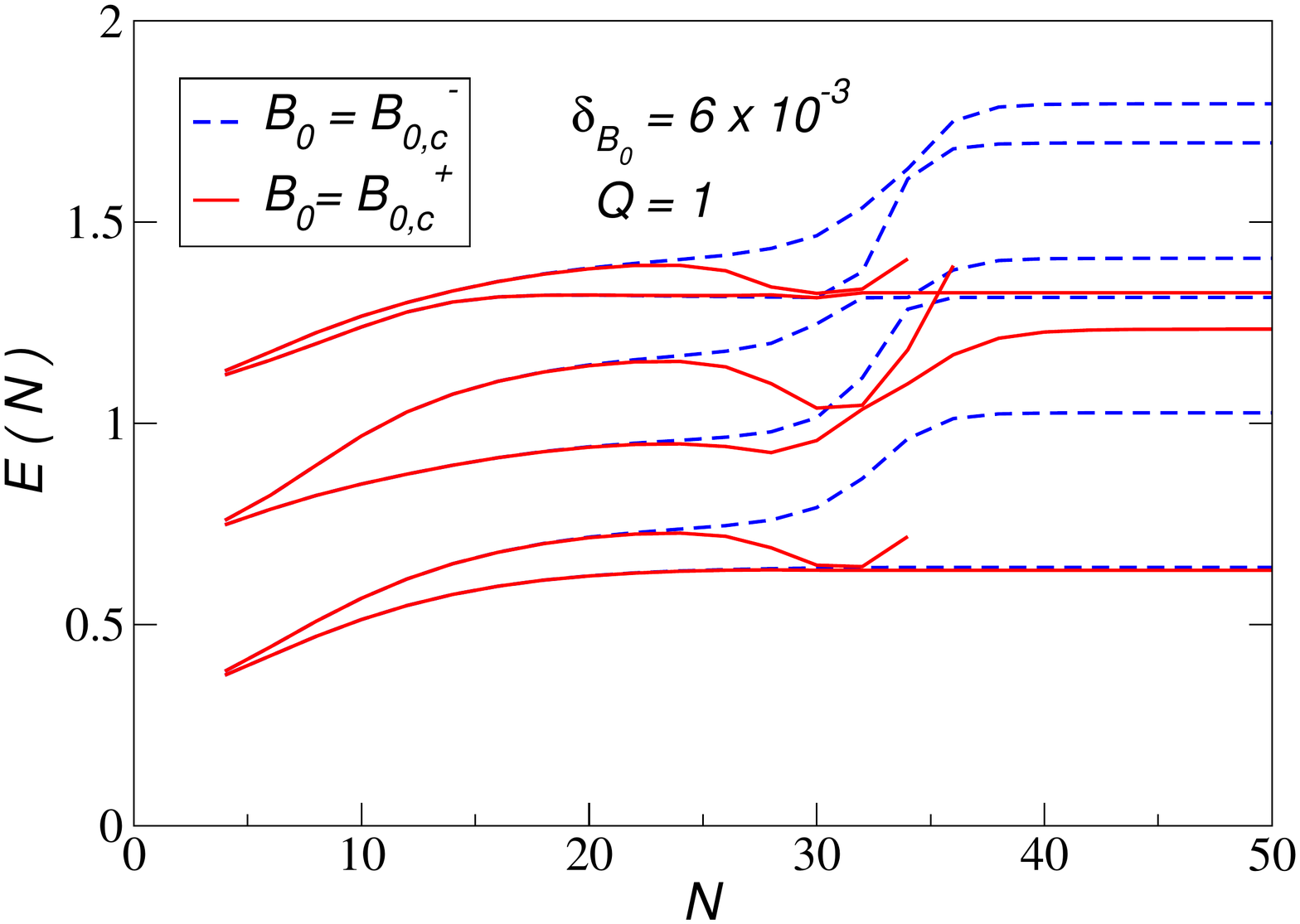}
\caption{Eigenenergy $E$ vs iteration number $N$ for the six lowest $Q=1$ states near the critical couplings for $\delta_{B_0}=6\times 10^{-3}$, showing flow away from the critical spectrum to the LM (solid) and K (dashed) sides.}
\label{fig9}
\end{figure}

The flows for the lowest six states closest to the critical point for ${\delta }_{B_0}=6\times{10}^{-3\ }$are shown in Fig.~\ref{fig9} for $Q=1$. 

\begin{figure}[!h]
\includegraphics[width=1.1\columnwidth]{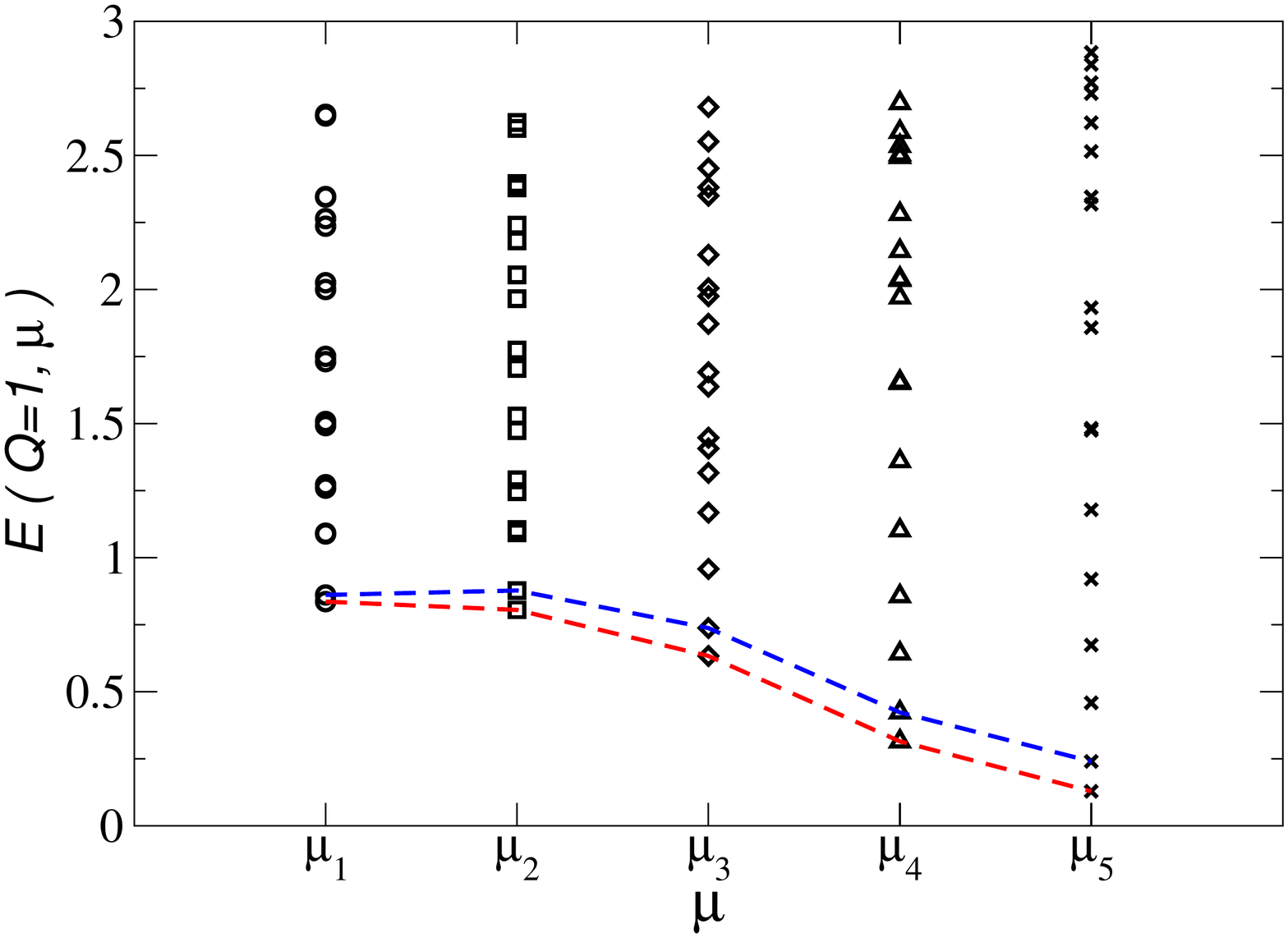}
\caption{Lowest $Q=1$ eigenvalues closest to the critical point ($N \simeq 24 ) $ for each case $\mu_1$ to $\mu_5$ defined in the caption of Fig. 8.
The dashed lines track the evolution of the splitting between the two lowest states.}
\label{fig10}
\end{figure}

The figure shows the flows for 
values of the coupling slightly below, and above the critical point. For $B_0<\ B_{0,c}$ the flows approach a plateau characteristic of the slow variation close to the unstable fixed point and eventually move away to the K fixed point. For $B_0>\ B_{0,c}$ 
we see the same critical values, which then move towards the LM fixed point. Notice that 
in this latter regime the states tend to become doubly degenerate and can be fitted 
to a spin-dependent potential 
scattering term corresponding to a finite ${\rho J}^*_z\ $term. This is also consistent 
with the full suppression of any tunneling term. Half of the flows tend to diverge quickly 
beyond $N \simeq 
34$ and are truncated in the figure. This lifting of the degeneracy is a 
purely numerical artifact in the LM regime and we do not expect it to have any significant 
bearing on our conclusions. In the critical regime,  $N \simeq
24 $ 
, the $Q=1$ states show 
a splitting due to an effective finite ${\Delta }^*\ $term which is absent in the $\Delta {\rm =0}$ 
BFKM case. Formally, a nonzero bare
$\Delta $\textbf{ }implies that the total spin projection 
along the z-axis is not conserved.  To see all this we show the lowest six 
estimated $Q=1$ critical eigenvalues ($N \simeq
 24$) for all ${\delta }_{B_0}$
trials in Fig.~\ref{fig10}.

 In  Figs.~\ref{fig8},~\ref{fig10}, and ~\ref{fig12}, $ \mu_1 $ corresponds to the critical couplings found for $ \delta_{B_0} = 6 \times 10^{-5} $,  $ \mu_2 $ to that for $ \delta_{B_0} = 6 \times 10^{-4} $ and so on.
Referring to the lowest two values in Fig.~\ref{fig10}, one can clearly see that the splitting increases continuously with ${\delta }_{B_0\ }$. For higher states, it becomes very difficult to follow the trend due to possible level crossings. 

\begin{figure}[!h]
\includegraphics[width=1.1\columnwidth]{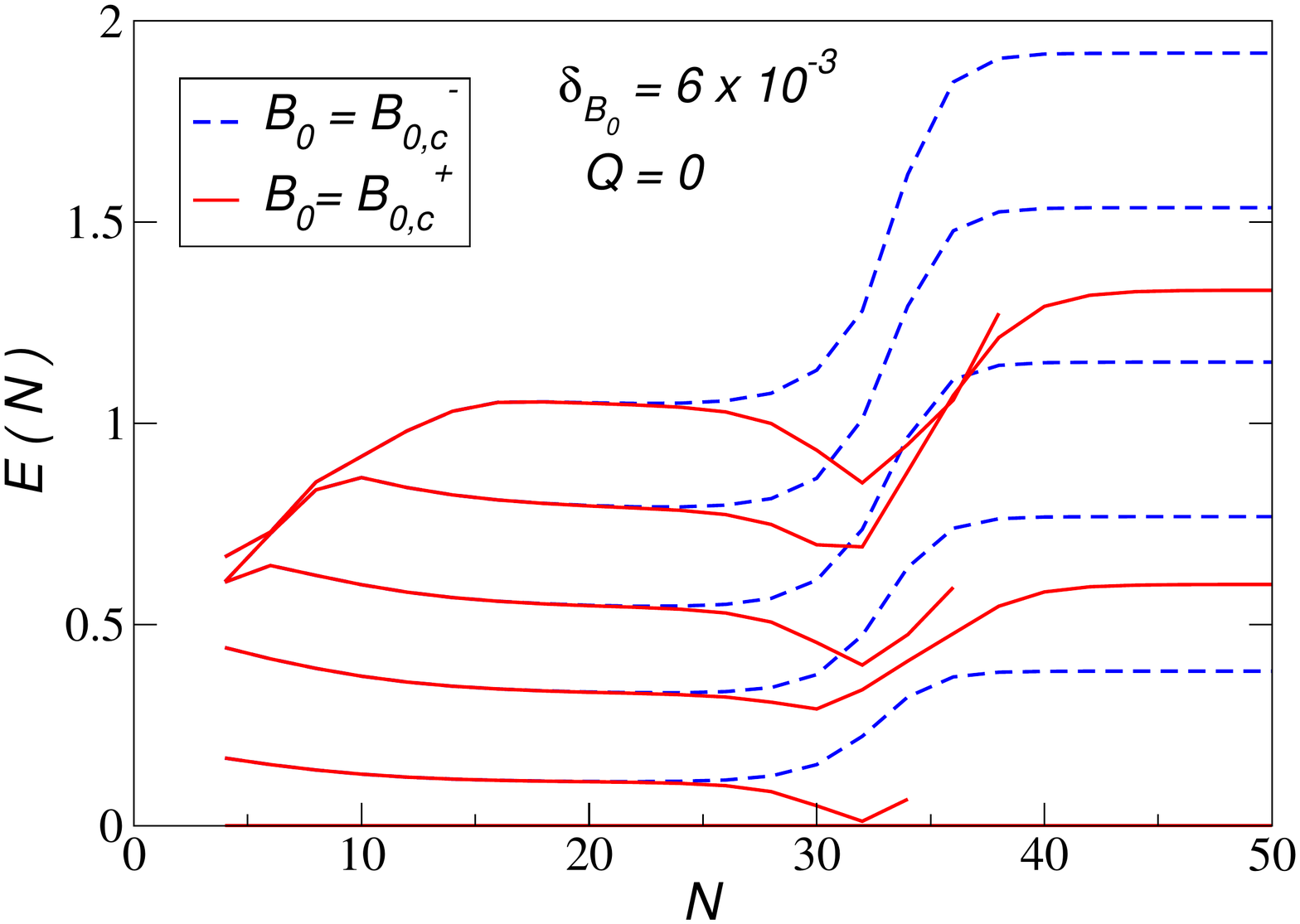}
\caption{Eigenenergy $E$ vs iteration number $N$ for the six lowest $Q=0$ states near the critical couplings for $\delta_{B_0}=6\times 10^{-3}$, showing flow away from the critical spectrum to the LM (solid) and
K (dashed) sides.}
\label{fig11}
\end{figure}

\begin{figure}[!h]
\includegraphics[width=1.1\columnwidth]{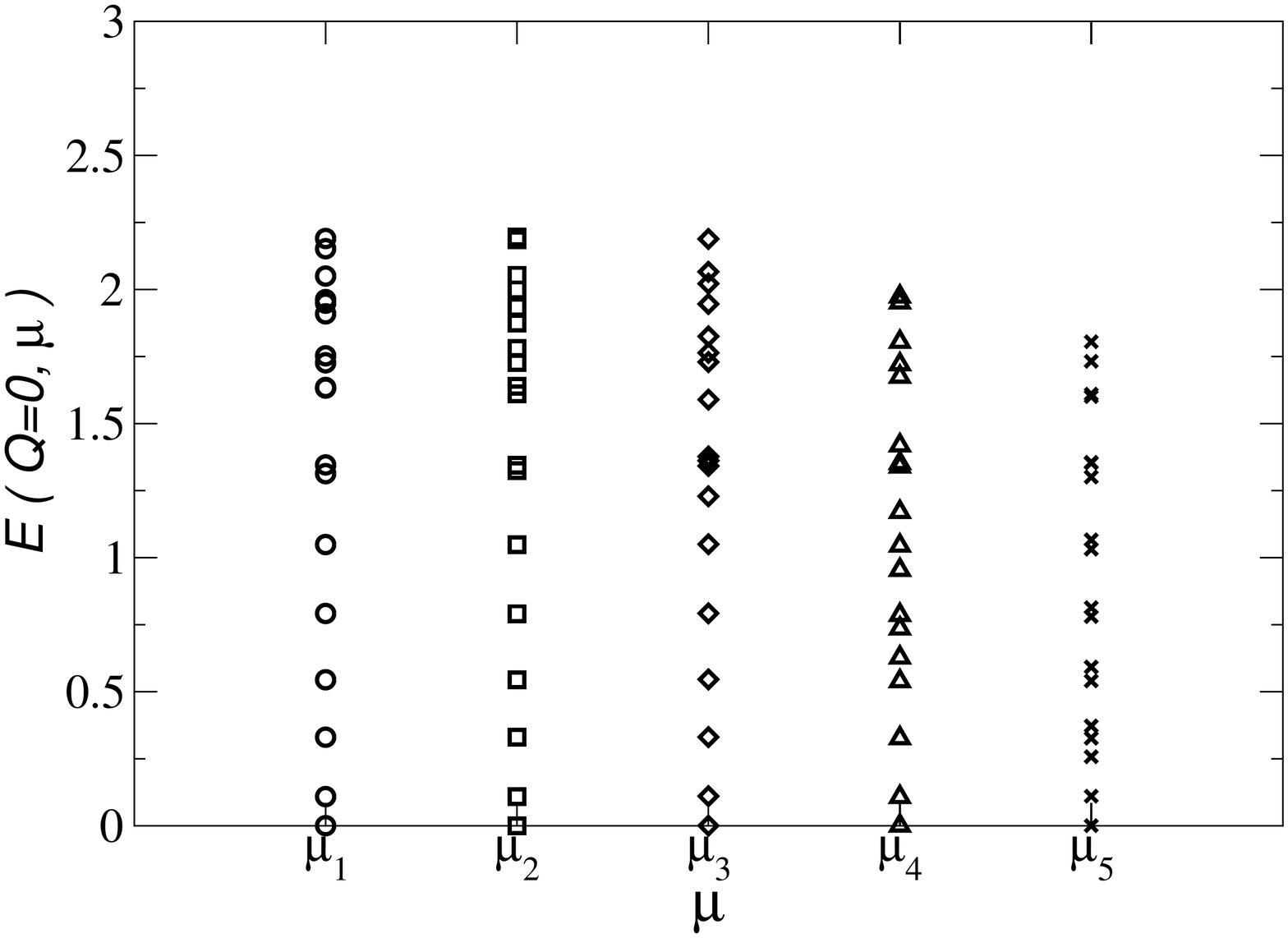}
\caption{Lowest $Q=0$ eigenvalues closest to the critical point ($N \simeq
24$) for each case $\mu_1$ to $\mu_5$ defined in
        the caption of Fig. 8.}
\label{fig12}
\end{figure}

\indent The $Q=0$ flows are shown in Fig.~\ref{fig11} for  $B_0$\textbf{ }around\textbf{ }$B_{0,c}$. It is well known \cite{ Weiss:1999, 41} that the truncation of the bosonic Hilbert space results in improperly converged boson eigenvalues at the LM fixed point. However, NRG truncation does not affect the critical bosonic spectrum. Figure~\ref{fig12} shows the estimates for the lowest few  $Q=0$ critical eigenvalues.  
One sees that some of these critical flows do not change with ${\delta }_{B_0}$. For the rest, it is more difficult to determine if the eigenvalues are indeed changing continuously as in the $Q=1$ sector. It is possible that the $ Q=0 $ eigenvalues which are changing are those associated with the fermionic excitation, while the constant values correspond to an unchanging purely bosonic sector.

\noindent 

\begin{figure}[h!]
\includegraphics[width=1.05\columnwidth]{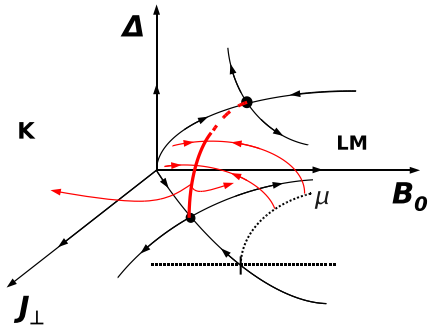}
\caption{Schematic flow diagram showing a line of unstable fixed points
extending from the $\Delta {\rm =0}$ BFKM critical fixed point with a conjectured
        extension (dashed) to the the SBM critical point lying on the plane
        $J_{\perp} = 0$.}
\label{fig13}
\end{figure}

 Although it is difficult at this stage to make strong statements 
 about the nature of these critical excitations we can attest at least to the 
 fact that in the $Q=1$ case they appear to change 
 continuously with ${\delta }_{B_0}$. A reasonable scenario is that the extra dissipation 
 provided by increased ${\delta }_{B_0}$ requires an increased renormalized 
 effective ${\Delta }^*$ at the critical point. Some evidence for this is also provided 
 since one requires an increasing bare ${\Delta }_c$\textbf{ }with increasing ${\delta }_{B_0}$
 in order to bring the flows closer to the critical surface. 
 The fact that the spectrum changes provides evidence that we are dealing with a line of 
 unstable
 fixed points extending from the $\Delta {\rm =0}$ BFKM\textbf{ }and having the same set of critical exponents as the former. In the alternate scenario alluded to above, a flow to a different unstable fixed point would produce essentially the same critical eigenvalues for all ${\delta }_{B_0}\ $cases. 
A schematic flow diagram for the 
transverse-field BFKM\textbf{ }showing a line of critical fixed points extending from the $\Delta {\rm =0}$ \textbf{ }case\textbf{ }is given in Fig.~\ref{fig13}.

We stress that the many-body spectrum of the unstable fixed point reflects the universal
properties of the critical point. 
Because of the non zero transverse field applied along the $x$ direction,
a magnetization will be generated that is directed along the $x$ direction,
as considered, {\it e.g.} in
Ref. \onlinecite{41}
in the SBM; this magnetization does not contain any critical singularity.
Nonetheless, the end result of tuning $\Delta$ is to make the many-body spectra 
of the unstable fixed points vary continuously, which is
captured by the line of 
fixed points shown in 
Fig.~\ref{fig13}.

We close this section by noting that 
the $ Q=1 $ eigenstates in the K phase do not recover the $ SU(2) $ symmetry characteristic of a simple Kondo singlet suggesting that there is a residual splitting 
due to the transverse field.  As shown in Fig.~\ref{fig9}, these splittings are larger in the converged 
 K phase eigenvalues than in the critical regime suggesting that the effective $ \Delta $ increases with flows away from the critical points.   
 At the same time, 
 the $Q = 1$ eigenvalues for $B_0 > B_{0,c}$  recover the two fold degeneracy typical of the LM fixed point with vanishing tunneling amplitude $\Delta^*$ in all of the $\delta_{B_0}$ cases considered.

\section{Conclusions}\label{Sec:Conclusions}

\noindent We have carried out numerical renormalization-group  studies of the Bose-Fermi Kondo model
in the presence of a transverse field $\Delta $ for different values of the coupling to a
 dissipative bosonic bath 
with a bath exponent $s{\rm =}{\rm 0.8}$. We found that the system can be tuned across 
a second-order quantum phase transition between a local-moment phase and a Kondo  screened phase. We also found that the transition is characterized by critical exponents identical to those of the $\Delta {\rm =0\ }$BFKM for $s{\rm =} {\rm 0.8}$, and exhibits
hyperscaling. A continuously varying critical spectrum suggests that these new fixed points are lying on a line of critical fixed points extending from the known $\Delta {\rm =0\ }$BFKM\textbf{ }critical point. 

Our results are interesting in their own right. Furthermore, in the Ising-anisotropic
Kondo lattice model, a transverse field introduces quantum fluctuations in the local-moment component. Through the extended dynamical mean field theory, a self-consistent Bose-Fermi Kondo model with Ising anisotropy and a transverse local field provides 
the means to study the Ising-anisotropic Kondo lattice model with a transverse field.
The results reported here will therefore have implications for the Kondo-destruction transitions
in the lattice model. 
In particular, our evidence for the line of unstable fixed points in the Bose-Fermi Kondo model
suggests a line of Kondo-destruction quantum critical point in the lattice model,
as in the proposed global phase diagram.
Concrete studies of the lattice model will be undertaken in the future.

\acknowledgements

We thank Stefan Kirchner, Jed Pixley and Jianda Wu for illuminating discussions.
This work has been in part supported by
NSF Grant Nos.\ DMR-1006985 and DMR-1107814, Robert A.\ Welch Foundation
Grant No.C-1411, and by the National Nuclear Security Administration of the U.S. DOE at LANL under Contract No.
DE-AC52-06NA25396 and the LANL LDRD Program. 

\appendix
\section{Accuracy in the determination of the critical exponents}
\label{Appendix:Accuracy of the results}

\begin{figure}[h!]
\includegraphics[width=1.1\columnwidth]{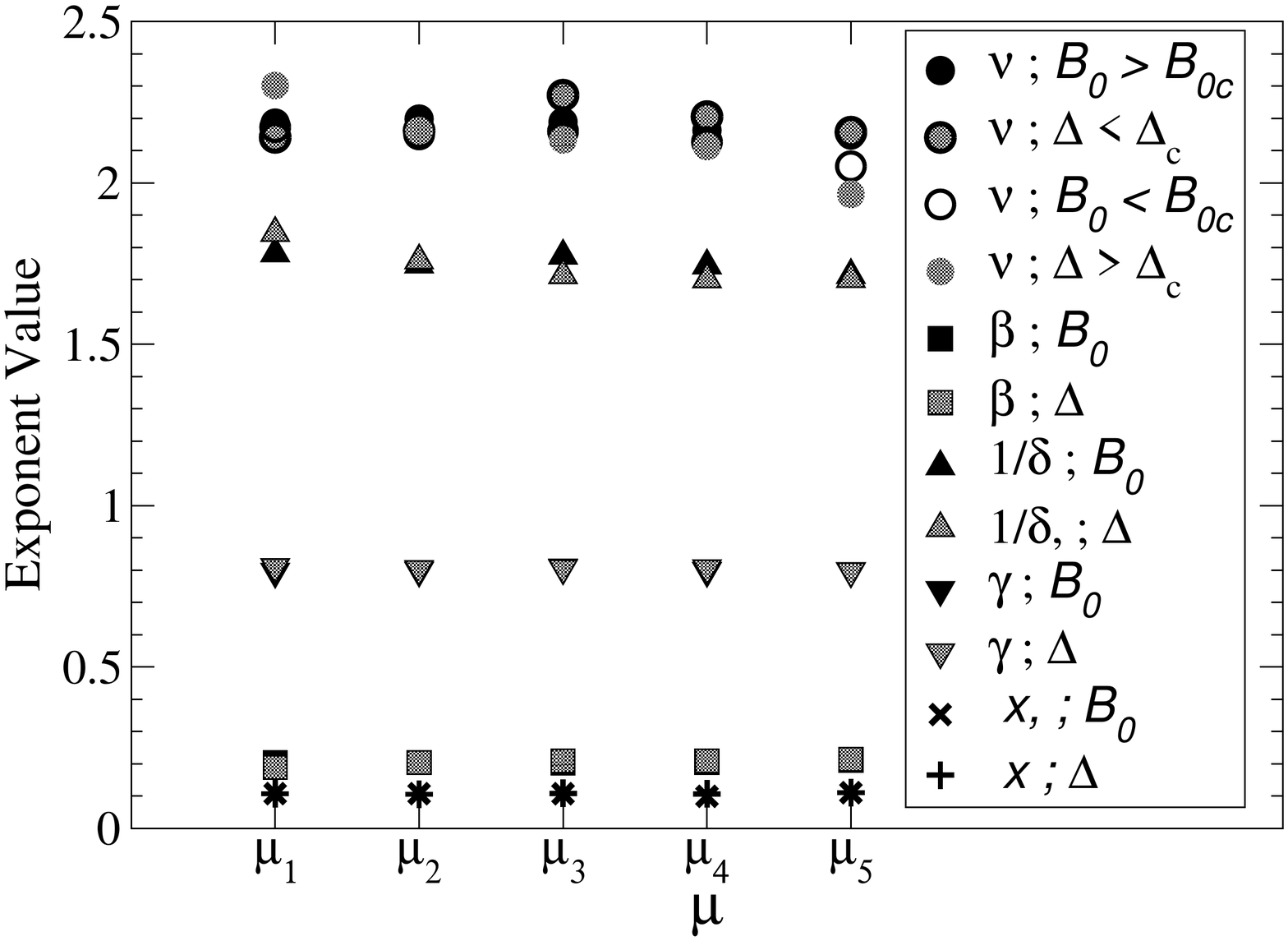}
\caption{Comparison of critical exponents obtained by tuning through the
        critical point by varying $B_0$ and by varying $\Delta$, showing
        results for each case $\mu_1$ to $\mu_5$ defined in the caption of
        Fig. 8.}
\label{figg14}
\end{figure}

 As a check on our estimates of critical
        exponents obtained by varying $B_0$ around $B_{0,c}$ at fixed
        $\Delta=\Delta_c$, we have also calculated these exponents by fixing
        $B_0=B_{0,c}$ and then tuning $\Delta$ through $\Delta_c$.
Figure~\ref{figg14}\textbf{ } compares the
        exponents obtained via these two methods, in each case retaining
        500 isospin multiplets in the NRG calculations.

There is some mismatch between the estimated exponents for the $\Delta $ and $B_0 $ tuning cases. This is especially pronounced for ${\nu }_{LM}$ and ${\nu }_K $ with the smallest and largest ${\delta }_{B_0} $  corresponding to cases $\mu_1 $ and $\mu_5 $,  respectively, in Fig.~\ref{figg14}. 

\indent 
The discrepancies in the first case may be attributable
to a vanishing $ \Delta^* $ at the unstable fixed point parameterized by $\mu_1 $.  For small changes in the bare $ \Delta $ close to ${\Delta }_c\ $ the change to the RG flow is particularly small due to proximity to the  critical point, making it difficult to
accurately determine the scaling property.

\indent The second, larger discrepancy in $\nu $  is noticed for $\mu_5 $. We observe in this case that the many-body eigenvalues approach the critical plateau at higher-numbered iterations and leave this vicinity sooner than all the other cases considered. This suggests that the flows do not come as close to an unstable fixed point as in the other cases.   

\indent Excluding these sources of error, it is clear that the exponents found by tuning $\Delta $ are the same as those found by tuning $B_0 $.

The error in the critical exponents 
is estimated to be of ${\rm O}\left({10}^{-1}\right)$ for $\nu $\, which was determined from the eigenvalue flows and of$\ {\rm O}({10}^{-2}$) for the thermodynamic critical exponents.

\end{document}